\documentclass[lettersize,journal]{IEEEtran}
\usepackage{amsmath,amsfonts}
\usepackage{cite}
\usepackage{array}
\usepackage{textcomp}
\usepackage{stfloats}
\usepackage{url}
\usepackage{setspace}
\usepackage{verbatim}
\usepackage{graphicx}  
\usepackage{subcaption}
\usepackage{textcomp}
\usepackage{xcolor}
\usepackage{multirow}
\usepackage{eurosym}
\usepackage{amssymb}
\usepackage{amsthm}
\usepackage[scr=rsfs]{mathalpha}
\usepackage{marginnote}
\usepackage[linesnumbered,ruled]{algorithm2e}
\usepackage{enumitem}
\usepackage{tablefootnote}
\usepackage{nomencl}
\makenomenclature
\DeclareMathOperator*{\argmin}{arg\,min}

\newcommand*{\Comb}[2]{{}^{#1}C_{#2}}%

\usepackage{gensymb}

\IEEEoverridecommandlockouts                              
\allowdisplaybreaks

\newtheorem{thm}{Theorem}
\newtheorem{remark}[thm]{Remark}
\newtheorem{definition}{Definition}
\newtheorem{proposition}{Proposition}

\usepackage{etoolbox}
\renewcommand\nomgroup[1]{%
  \item[\bfseries
  \ifstrequal{#1}{P}{\textit{C. Variables}}{%
  \ifstrequal{#1}{N}{\textit{A. Sets}}{%
  \ifstrequal{#1}{O}{\textit{B. Parameters}}{}}}%
]}

\begin{document}
	
	\title{A User-centric Game for Balancing V2G Benefits with Battery Degradation of Electric Vehicles } %The Name of the Title is Yet to be Decided
 
    \author{Arghya Mallick,~\IEEEmembership{Student Member,~IEEE,} Georgios Pantazis, \\
    Peyman Mohajerin Esfahani, Sergio Grammatico,~\IEEEmembership{Senior Member,~IEEE}  

    \thanks{ 
    This work is supported by the European Union (EU) under the project Drive2X (grant number: 101056934). The authors would like to acknowledge the valuable suggestions by Prof. Gautham Ram Chandra Mouli during the preparation of this manuscript. This article is accepted for publication in \textit{IEEE Transactions on Transportation Electrification} (\textit{Corresponding author: Arghya Mallick)}.
    
    All the authors are with the Delft Center for Systems and Control, TU Delft, Mekelweg 5, 2628 CN Delft, Netherlands. Emails: \{A.Mallick, G.Pantazis, P.MohajerinEsfahani, S.Grammatico\}{@}tudelft.nl}} % <-this % stops a space

% The paper headers
\markboth{}%
{Shell \MakeLowercase{\textit{et al.}}: A Sample Article Using IEEEtran.cls for IEEE Journals}

\maketitle

\begin{abstract}
We present a novel user-centric vehicle-to-grid (V2G) framework that enables electric vehicle (EV) users to balance the trade-off between financial benefits from V2G and battery health degradation based on individual preference signals. Specifically, we introduce a game-theoretic model that treats the conflicting objectives of maximizing revenue from V2G participation and minimizing battery health degradation as two self-interested players. Via an enhanced semi-empirical battery health degradation model, we propose a finite-horizon smart charging strategy based on a horizon-splitting approach. Our method determines an appropriate allocation of time slots to each player according to the user’s preferences, allowing for a flexible, personalized trade-off between V2G revenue and battery longevity. We conduct a comparative study between our approach and a multi-objective optimization formulation by evaluating the robustness of the charging schedules under parameter uncertainty and providing empirical estimates of regret and sensitivity. We validate our approach using realistic datasets through extensive trade-off studies that explore the impact of factors such as ambient temperature, charger type, and battery capacity, offering key insights to guide EV users in making informed decisions about V2G participation.
\end{abstract}
	

    %\nomenclature[P]{\(P,Q\)}{Active and reactive power flow on lines.}    
    %\printnomenclature
\section{Introduction}
\IEEEPARstart{T}{he} global energy sector is currently being transformed by the rapid expansion of renewable energy sources and the widespread adoption of electric vehicles (EVs), driving research on the impact of those technologies to the electricity grid and the integration of energy storage systems to address emerging challenges. The energy storage industry has historically relied heavily on lithium-ion batteries (LIBs) \cite{ziegler2021re}; however, due to factors such as material security and other concerns \cite{material_security}, reducing the reliance on new battery purchases has become essential. Along this direction, researchers have proposed repurposing existing EV batteries for grid-support services \cite{liu2013opportunities,aguilar2024potential,xu2023electric}. This idea, known as vehicle-to-grid (V2G), was first introduced in \cite{kempton1997electric} and has been tested in over 100 pilot projects worldwide since 2002 \cite{brooks2002vehicle,v2g_uk}. Despite these efforts, widespread adoption of V2G technology has been slow, primarily due to regulatory challenges and societal resistance \cite{aguilar2024potential}. The key to driving regulatory reform lies in gaining the voluntary participation of EV users in V2G programs. A crucial qualitative study in \cite{van2021factors} identified the main reasons for societal hesitation towards V2G, with two prominent concerns being the uncertainty of financial benefits and the potential for battery degradation. To illustrate these concerns, one EV user in \cite{van2021factors} stated, \textit{``If discharging for V2G-mode is done only a couple of times per year, then I would find it acceptable to participate in V2G. But if you do V2G on a daily basis (hundreds of times per year), I believe that the battery pack will be damaged and then I would not participate.”} Given these concerns, there is a pressing need for robust research to evaluate the effects of V2G participation on battery health and the associated financial benefits for EV owners.  

Few studies \cite{bishop2013evaluating,wang2016quantifying,thingvad2021empirical,razi2023predictive,lu2024coordinated} have rigorously investigated the impact of V2G services on EV battery health. In \cite{bishop2013evaluating}, the authors demonstrate that battery degradation is accelerated when providing bulk energy and ancillary services through V2G. However, their degradation model does not account for the distinction between battery cell temperature and ambient temperature, potentially leading to inaccurate results. Similarly, the study in \cite{wang2016quantifying} assesses battery health degradation due to V2G participation, concluding that the degradation is negligible when compared to naturally occurring factors such as driving and calendar aging. Nevertheless, neither the study in \cite{bishop2013evaluating} nor that in \cite{wang2016quantifying} offers a comprehensive analysis of the trade-offs between the financial benefits of V2G participation and the associated financial losses due to battery degradation. Furthermore, these studies do not consider the effects of time-varying charging and discharging power (commonly referred to as smart charging) on battery health during V2G sessions—a critical omission given that a recent study \cite{brinkel2024enhancing} has demonstrated the financial benefits of implementing smart charging for V2G services. Authors in \cite{thingvad2021empirical} present an experimental study involving real-world EVs engaged in V2G services over a five-year period. The authors propose an empirical method for measuring battery capacity and demonstrate a total capacity fade of $17.8 \%$, with one-third of this degradation attributable to cyclic processes, including V2G usage and daily driving. Finally, studies such as \cite{razi2023predictive,lu2024coordinated} have aimed to incorporate user-centric considerations in developing smart-charging algorithms under V2G services, taking battery degradation into account. However, these works do not provide a framework that enables users to individually balance their interests between maximizing V2G participation and minimizing battery degradation.  

While the above studies provide insights into EV battery health degradation from V2G participation, they fall short of offering solid recommendations for EV owners regarding the optimal timing and extent of participation in V2G to achieve favorable outcomes. To address this gap, we propose a novel framework that captures the trade-offs between two conflicting objectives: (a) the financial gains from V2G participation and (b) the degradation of battery health. 
For the first time, we model this trade-off using game theory to simulate the inherent competition between these objectives. In our game-theoretic framework, each objective is assigned to a player, and the players engage in a strategic game constrained by factors such as EV charger ratings, required energy levels at the end of V2G sessions, and battery dynamics. Building on the game equilibrium solution, we delve deeper into practical scenarios, examining the effects of varying ambient temperatures, EV charger types, battery capacities, and other relevant factors. Finally, we present key insights that hold significant potential value for the stakeholders.            

Our main contributions with respect to the related literature are the following: 
\begin{enumerate}
    \item
    \textbf{Novel equilibrium concept:} We propose a game-theoretic trade-off between the two conflicting objectives of V2G exploitation and battery degradation (Section III) where the two players involved determine the level of participation via an a-priori parameter while sharing common constraints. The key difference in contrast with the existing approaches \cite{razi2023predictive,lu2024coordinated} is a user-defined hyperparameter allowing users to balance their level of V2G participation and battery degradation, facilitating a more user-friendly smart charging solution for widespread EV adoption.  
    \item \textbf{Complexity vs accuracy:} For the battery degradation modeling, we provide a balance between complexity and accuracy by deriving a smooth approximation of the empirical battery health degradation model (Section II) proposed in \cite{wang2014degradation}, thus making it suitable for integration into an optimization framework. Unlike previous studies \cite{thingvad2021empirical,razi2023predictive}, \cite{lu2024coordinated}, we calculate an offline solution for the battery temperature dynamics, which is incorporated directly into the semi-empirical degradation model, enhancing its accuracy.
    \item \textbf{Sensitivity and trade-off analysis:} We compare the proposed framework with a multi-objective optimization approach and a state-of-the-art MPC-based method \cite{lu2024coordinated} using sensitivity and regret metrics, and the empirical results indicate that our method is significantly more robust when the parameters of the objective function are subject to perturbations. We perform a comprehensive trade-off analysis (Section IV) of our methodology to assess the impact of factors such as ambient temperature, EV charger ratings, V2G tariff structures, and battery capacities. This analysis aims to alleviate uncertainties for EV users, encouraging more active participation in V2G programs.  
\end{enumerate}

Finally, Section V concludes the paper by proposing directions for further research.

\section{Modelling of battery health degradation}
 To accurately represent the health degradation of a Li-ion battery, we leverage a capacity fading model from \cite{wang2014degradation}, which explicitly relates the capacity fading of the Li-ion battery with the so-called operating C-rate of the battery. The operating C-rate of a battery is an important unit used for measuring how fast a battery is charged or discharged with respect to its capacity. 
 The proposed model is based on the distinction between two primary forms of capacity fading, namely calendar aging and cyclic aging. 
 Previous works have employed physics-based models \cite{safari2008multimodal} and machine learning models \cite{zhang2020identifying,severson2019data} to predict battery degradation. However, both approaches have limitations, including high computational complexity and reliance on extensive experimental datasets. To offer a balance between model complexity and accuracy and thus mitigate the challenges encountered in \cite{safari2008multimodal,zhang2020identifying,severson2019data},  our work follows a semi-empirical approach, integrating battery temperature dynamics represented by differential equations with empirical models on calendar and cyclic aging. This leads to a model which is fit for fast and reliable real-time operation with reasonable accuracy. %To strike a balance between model complexity and accuracy, we opt for semi-empirical battery degradation models.
\subsection{Calendar and cyclic aging}
Calendar aging refers to capacity loss due to an irreversible process of gradual self-discharge. This capacity loss is the result of lithium inventory loss during the solid-state-interphase (SEI)
formation at the graphite negative electrode \cite{wang2014degradation}. In other words, the growth of the SEI layer consumes lithium, causing irreversible capacity loss. 
The capacity loss in calendar aging increases with time ($t$) and the temperature ($T_b$) of the battery. In this work, we leverage an empirical model developed in \cite{smith2012comparison,wang2014degradation} for capturing calendar aging of Sanyo UR18650W Li-ion battery cells. Note that our analysis and methodology are not restricted to this particular battery model and can, in fact, be appropriately adapted to handle other Li-ion models. According to this model, the capacity loss due to calendar aging $Q^{\text{cal}}_{\text{loss},\%}$ is given by 
\begin{equation} \label{calendar_eq}
    Q^{\text{cal}}_{\text{loss},\%} = A \exp{\left(-\frac{E_a}{RT_{b}}\right)} \sqrt{t},
\end{equation}
 where $A$ is a pre-exponential factor, $E_a$ is the activation energy, which is $24.5$ kJ/mol, and $R$ is the ideal gas constant, which is $8.314$ J/mol/K. Note that the temperature $T_{b}$ is implicitly also a function of time.
 
To model cyclic aging, we follow the empirical model proposed in \cite{wang2014degradation}: 
\begin{subequations}\label{emp_model}
    \begin{align}
        \label{q_l}
        & Q_{\text{loss},\%}^{\text{cyc}} = B_{1} (\exp{(B_{2}|I_{\text{rate}}|)}C_{\text{rated}}n_{\text{cycle}}), \\
        \label{B_1}
        & B_{1}= aT_{b}^2 + bT_{b} + c, \\
        & B_{2}= dT_{b} + e,
    \end{align}
\end{subequations}
where $C_{\text{rated}}$ is the capacity of the battery in $Ah$, $T_{b}$ is the battery temperature in $K$, $n_{\text{cycle}}$ is the number of cycles, $I_{\text{rate}}$ is the operating C-rate of battery, and $a,b,c,d,e$ are constants calculated from the experimental data and given in Table \ref{table_bat_model}. Equation \eqref{emp_model} is the degradation model of a single battery cell. We need to scale down the battery pack level charging/discharging power to a single-cell level. To alleviate computational challenges, while retaining an accurate model, we consider a smooth approximation of Equation \eqref{q_l}: 
\begin{equation} \label{loss_model}
    Q_{\text{loss},t}^{\text{cyc}}= B_{1,t}C_{\text{rated}}^2 \hat{n} \left(1+\frac{\hat{B}_{2,t}^2P_{\text{bat},t}^2}{hs^2} \right),
\end{equation}
where $\hat{n}= \frac{n_{\text{max}}\Delta t}{T\times100}$, $P_{\text{bat},t}$ is the charging/discharging power of the battery (positive value implies charging and negative value indicates discharging), $n_{\text{max}}$ is the maximum number of full cycles\footnote{Assuming $t_f$ being the required time for a complete charging to discharging cycle, $n_{\text{max}}$ is equal to $\lceil T/t_f \rceil)$ where, $\lceil \cdot \rceil$ is the least integer function.} by time $T$, $\hat{B}_{2,t}= \frac{B_{2,t}}{V_{\text{bat}}C_{\text{rated}}}$ with the coefficient $\frac{1}{V_{\text{bat}}C_{\text{rated}}}$ being used to convert $I_{\text{rate}}$ to $P_{\text{bat},t}$, $s=n_{\text{series}}\times n_{\text{parallel}}$ is the scaling factor for bringing down $P_{\text{bat},t}$ to a single cell level, $n_{\text{series}}$ and $n_{\text{parallel}}$ are respectively the number of series connected battery cells and parallel connected strings of cells forming the battery in the EV. $V_{\text{bat}}$ is the terminal voltage of a single battery cell, and $h$ is an appropriate curve fitting parameter for approximation of \eqref{emp_model}. Note that, $B_{1,t}$, and $\hat{B}_{2,t}$ are time varrying due to $T_b$. The detailed derivation of \eqref{loss_model} is given in Appendix A. 
\begin{table}
    \centering
    \caption{Experimental parameters of empirical model \cite{wang2014degradation}.}
    \begin{tabular}{p{1.6 cm} p{3.6 cm}}
         \hline
         \textbf{Parameters}  & \textbf{Value} \\
         \hline
         $a$ & $8.61\times 10^{-6}$,1/Ah-K2  \\
         $b$ & $-5.13 \times 10^{-3}$ 1/Ah-K \\
         $c$ & $7.63 \times 10^{-1}$ 1/Ah  \\
         $d$ & $-6.7 \times 10^{-3}$ 1/K-(C-rate) \\
         $e$ & $2.35$ 1/(C-rate)  \\
         $A$ & $14,867$ 1/day$^{0.5}$ \\ 
         $h$ & $0.0465$ \\ 
         \hline
    \end{tabular}
     \label{table_bat_model}
\end{table}
\begin{remark}
    We note that the degradation result of \eqref{calendar_eq} and \eqref{emp_model} have recently been compared with experimental results in \cite{thingvad2021empirical} (see Fig. $10$ and $11$) and found to be reasonably accurate. Moreover, authors in \cite{thingvad2021empirical} conducted the comparison considering the typical usage of EVs under different V2G services. Note that our recent work \cite{mallick2025user} provides a more accurate data-driven battery degradation model, which could easily be integrated into our proposed framework due to its convexity property with respect to charging/discharging power.
\end{remark}

In the model above, a crucial factor is the accuracy of the battery temperature $T_b$, which is not straightforward to evaluate. To account for this, we combine our battery degradation model with a variant of the dynamic battery temperature model proposed in \cite{wang2016quantifying}. 
\subsection{Dynamic battery temperature model}
The temperature of the battery ($T_b$) cell is influenced by several factors, including ambient temperature, energy consumed by the HVAC system of the car, charging/discharging rate,  and efficiency of battery thermal management systems. Authors in \cite{neubauer2014thru} have utilized a car-level lumped capacitance thermal network approach for modeling $T_b$ considering the above factors. In \cite{wang2016quantifying}, authors have used the model of \cite{neubauer2014thru} to develop the following coupled differential equations for $T_b$. 
\begin{subequations} \label{temp_model}
    \begin{align}
        & M_c\dot{T_c} = K_{ac}(T_a-T_c) + K_{bc}(T_b-T_c) + q_{\text{rad}} + q_{\text{hvac}}, \\
        & M_b\dot{T_b} = K_{ab}(T_a-T_b) + K_{bc}(T_c-T_b) + q_{\text{btms}} + Q, 
    \end{align}
\end{subequations}
where $T_a$ is the ambient temperature, $T_c$ is the temperature of the cabin of the EV, $K_{ab}$ and $K_{ac}$ are, respectively, the effective heat transfer coefficients between ambiance and battery, and ambiance and cabin; $M_c$ and $M_b$ are the thermal mass of vehicle cabin and battery respectively; $q_{\text{rad}}$ is the solar radiance falling directly on the car, $q_{\text{hvac}}$ is the heat added to or, removed from the cabin by HVAC (Heating, ventilation, and air conditioning) system of the EV, $q_{\text{btms}}$ is the heat added to or, removed from battery pack by the battery thermal management system, and finally $Q$ represents the heat generated from the battery during charging/discharging process. In particular, $Q$ is equal to $I^2R_{\text{int}}$ where $I$ is the charging/discharging current, and $R_{\text{int}}$ is an estimated internal resistance of the battery. To obtain a solution of \eqref{temp_model}, we assume for simplicity the following: (1) $q_{\text{rad}}$ is zero, which implies that the EV is in the shade of a car parking lot or charging station; (2) $q_{\text{hvac}}$ is also zero as the HVAC system is most of the time turned off during parking; (3) the goal of the battery thermal management system is to regulate the battery temperature $T_b$, when the ambient temperature $T_a$ exceeds the $15^o-30^o C$ range\cite{lin2021review}. In case $T_a$ is between $15^o-30^o C$, $q_{\text{btms}}$ can be roughly assumed to be equal to $-0.9Q$ considering $90 \%$ heat removal efficiency of battery thermal management system. We impose the above assumptions as it is difficult to provide a good estimate of $q_{\text{rad}}$, $q_{\text{hvac}}$, and $q_{\text{btms}}$ without having access to a detailed vehicle model and sensory data. Furthermore, these parameters are inherently scenario-specific and may vary depending on environmental conditions and vehicle-specific characteristics. Ideally, an adaptive mechanism should be integrated into the proposed framework to dynamically update it over time. However, estimates of $q_{\text{hvac}}$ can be obtained by measuring the energy consumption of the HVAC system. To access $q_{\text{rad}}$ for a particular region, open-source websites such as NASA Power \cite{rad_data} could be used. Once the values of $q_{\text{rad}}$, $q_{\text{hvac}}$, and $q_{\text{btms}}$ are retrieved, they can be incorporated in the solution of \eqref{temp_model}.

\begin{figure}
    \centering
    \includegraphics[width=0.85\linewidth]{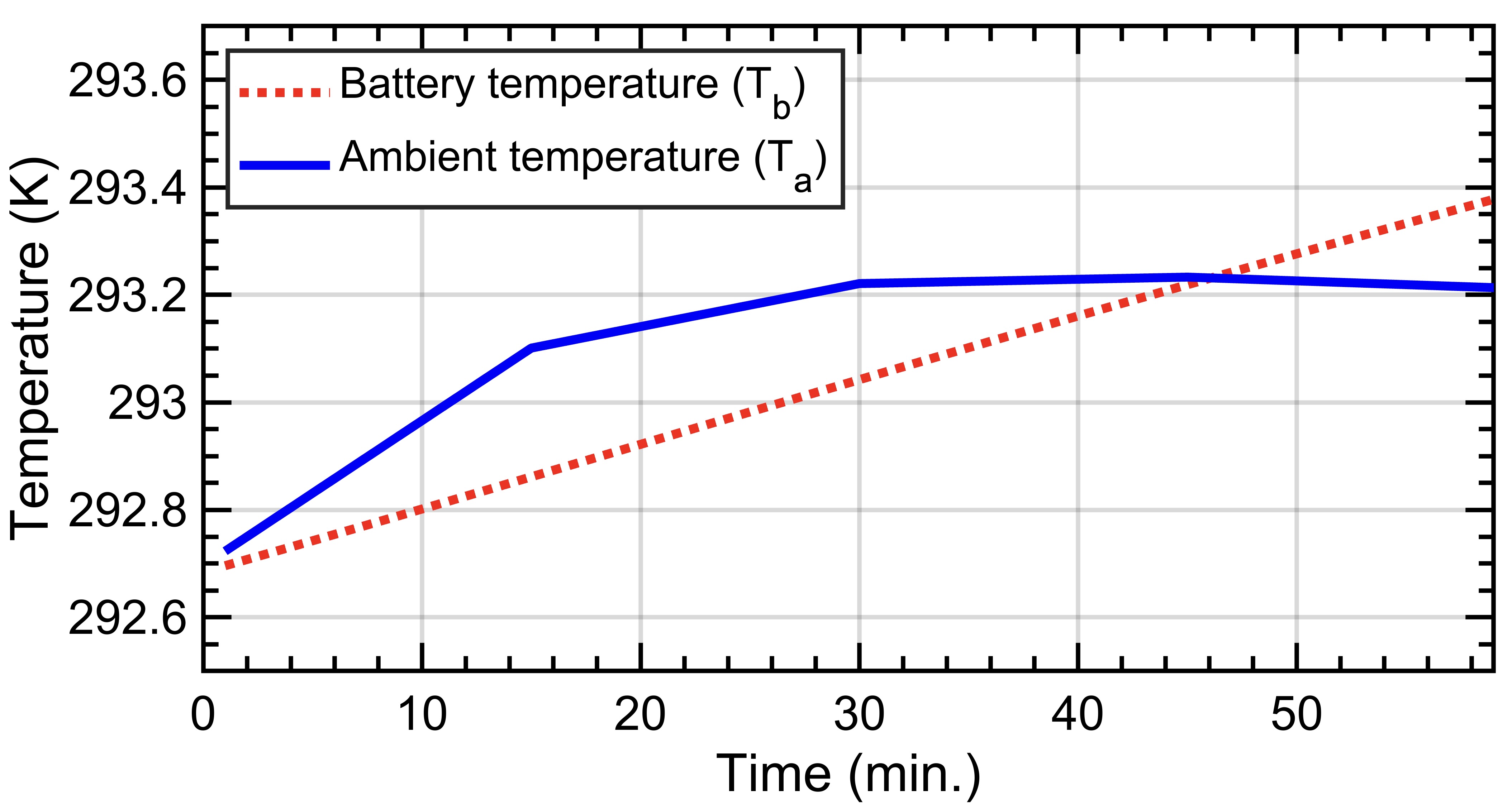}
    \caption{\small Evolution of battery temperature ($T_b$) with respect to ambient temperature ($T_a$).}
    \label{bat_temp_plot}
\end{figure}

In our proposed formulation, we use the battery temperature $T_b$ as a parameter. However, note that $T_b$ is itself a function of the battery's current $I$, which plays the role of the decision variable in the subsequent developments. Thus, for the heat $Q$ generated from the battery's operation, we use its approximation $\hat{I}^2R_{\text{int}}$ where, $\hat{I}:=\rho (\frac{P_{\text{max}}}{V_{\text{bat}}})$, with $P_{\text{max}}$ being the maximum charging/discharging power, and $\rho$ is a user-defined hyperparameter for denoting V2G participation level. The role of the hyperparameter $\rho$ is further elaborated in Section III.B. By considering values of other parameters in \eqref{temp_model} as given in Table 1 of \cite{neubauer2014thru}, we explicitly solve the dynamical equations in \eqref{temp_model}, where the input is the profile of $T_a$. Thus, we plot $T_b$ in Fig. \ref{bat_temp_plot} for a one-hour duration and find the gradual difference it creates with $T_a$.  
We use this pre-calculated profile of $T_b$ in \eqref{emp_model} for better accuracy.   
\section{ A game-theoretic approach to balance V2G participation level with battery degradation }  
\subsection{Preliminaries on game theory}
To model the trade-off between the V2G exploitation and battery degradation, we consider a non-cooperative game between $N=2$ players, indexed by $i \in \mathcal{I}:=\{1,2\}$.  Each player $i$ makes decisions $u^i \in \mathbb{R}^{n_i}$ in the feasibility set $\Omega_{i}(u^{-i}) \subseteq \mathbb{R}^{n_i}$ such that, the objective function (or, utility function) $\theta^i : \mathbb{R}^n \rightarrow \mathbb{R}$ is minimized, where $n=\sum_{i =1}^N n_i$, and $u^{-i}:= \text{col}((u^j)_{j\in \mathcal{I}\setminus \{i\}})$. Each player's strategy depends on the other players' strategies $u^{-i}$. Finally, let $\mathbf{u}:= \text{col}(u^1,u^2) \in \mathbb{R}^{2n}$ be the collection of the decision variables of both players, and $\Omega(\mathbf{u}):= \prod_{i=1}^{N} \Omega_i(u^{-i})$ be the collective feasible set. We define game $\mathscr{G}$ as the two coupled optimization problems as follows: 

\begin{equation} \label{gnep}
\mathscr{G}: \quad \forall \: i\in \mathcal{I}: 
     \begin{cases}
        \min\limits_{u^i} \quad & \theta^{i}(u^{i},u^{-i}), \\
        \  \text{s.t.} \quad & u^{i} \in \Omega_i (u^{-i}).
    \end{cases}
\end{equation}
Let us focus on the solution concept of generalized Nash equilibrium (GNE). 
\begin{definition} (Generalized Nash equilibrium \cite{facchinei2010generalized})
    A generalized Nash equilibrium  of the game $\mathscr{G}$ in (\ref{gnep}) is a tuple of strategies $\overline{\mathbf{u}}:= \text{col}(\overline{u}^1,\overline{u}^2) \in \Omega$ such that, for  each player $i \in \mathcal{I}$, we have 
    \begin{align*}
       \theta^i(\overline{u}^i,\overline{u}^{-i}) \leq \theta^i(y^i,\overline{u}^{-i}), \quad \forall y^i \in \Omega_i(\overline{u}^{-i}). 
    \end{align*}
\end{definition} 
 Therefore, at a GNE, no player can improve their objective function value by unilaterally changing to any other feasible solution, given the other player's decision. We note that thanks to the convexity of $\theta^{i}(.,u^{-i})$ for all $i \in \mathcal{I}$ and the compactness of the feasible set, the GNE problem in  (\ref{gnep}) admits at least one solution \cite{facchinei2010generalized}. 

To conclude the subsection, we recall the definition of a special class of games that behave similarly to a single optimization problem: 

\begin{definition} \label{def:potential}(Exact potential games  \cite{facchinei2011decomposition})  
Consider $\mathscr{G}$ on a non-empty, closed set  $\Omega = \prod_{i=1}^N \Omega_i(u^{-i})$, where  $\Omega_i(u^{-i}) := \{u^i \in D_i : (u^i,u^{-i}) \in \Omega \}$, and $D_i$ a closed constraint set local to agent $i$.  $\mathscr{G}$ is an exact generalized potential game if there exists a continuous function $P(\mathbf{u}): \mathbb{R}^n \rightarrow \mathbb{R}$ such that the following holds: 
    \begin{equation}
        P(x^i,u^{-i}) - P(y^i,u^{-i}) = \theta^i(x^i,u^{-i})- \theta^i(y^i,u^{-i}) , \nonumber 
    \end{equation}
for all players $i \in \mathcal{I} $, and all strategies $x^i,y^i \in \Omega_i(u^{-i})$. 
        \hfill $\square$
\end{definition}
Definition \ref{def:potential} implies that each player's seemingly selfish behaviour is aligned with an underlying common goal, represented by the potential function $P$.
\subsection{Horizon-splitting method for V2G smart charging}
 Our model considers an EV parking lot or charging station, also referred to as a system operator, which provides V2G services to the parked EVs. Specifically, the system operator delivers grid support services through the solution of an optimization problem, ensuring the satisfaction of operational constraints and charging options of the parked EVs. The operator determines the terms of energy exchange with the parked EVs by issuing a V2G price signal to each incoming EV \cite{lai2022pricing}. 

As opposed to previous methodologies, our work takes the EV owners' perspective and develops a framework that allows them to individually choose their participation level in V2G services. In our model, this personalized participation level is controlled via a hyperparameter that illustrates the subjective importance each EV owner assigns to their EV's possible battery degradation because of charging/discharging during V2G. Without loss of generality, we consider that each vehicle is charging/discharging for $T$ time intervals, each with duration of $\Delta t=15$ minutes. Our aim is to devise a smart charging/discharging strategy for each time interval $t \in [T]$, where $[T]:= \{1,2, \dots T\}$. To model the inherent competition between V2G services and battery degradation (BD), we consider a game between a player that optimizes with respect to the V2G revenue and a player that minimizes their battery degradation cost.

Given this game-theoretic model, each vehicle user determines the value of a hyperparameter that illustrates how much they value their battery health. Note that being overly conservative with respect to their battery degradation might be less beneficial, especially when V2G yields significantly higher returns to the EV user.  This hyperparameter is then used to determine the trade-off between V2G and BD by assigning a subset of $[T]$ to one player, and the rest intervals to the other player. Each of the players optimizes their respective objectives during their allocated intervals only. 

Let us define $\mathcal{T}_m^w$ as the set of possible subsets of $[T]$, where $w$ indicates the number of intervals chosen and $m$ indicates which particular subset of intervals is chosen from $[T]$. 

 We define the decision vector for player 1 (V2G) as $u^1:= \text{col}((P_{\text{bat},t})_{t \in \mathcal{T}_m^w})$, where $P_{\text{bat},t}>0 \ (<0)$ signifies the charging (discharging) of the EV's battery; the decision vector for player 2 (BD) is defined as $u^2 := \text{col}((P_{\text{bat},t})_{t \in \Tilde{\mathcal{T}}_m^w})$, where $\Tilde{\mathcal{T}}_m^w:=  [T]\setminus \mathcal{T}_m^w$. The way the horizon is split encapsulates the relative importance that the EV user assigns between the two objectives.
 
We first introduce the operational constraints that need to be satisfied at each time step $t \in [T]$. The bounds on charging/discharging power of  EV charger are as follows: 
    \begin{equation} \label{p_bounds}
        \underline{P} \leq P_{\text{bat},t} \leq \overline{P}, \: \forall t \in \mathcal{T}_m^w \: \text{or} \: \Tilde{\mathcal{T}}_m^w .
    \end{equation}
    The constraints related to battery energy dynamics are given by
    \begin{equation} \label{e_bounds}
        \underline{E} \leq E_0 + \eta_{\text{avg}}\Delta t\left(\sum_{j=1}^{t} P_{\text{bat},j} \right) \leq \overline{E}, \quad \forall \: t \in \mathcal{T}_m^w \: \text{or} \: \Tilde{\mathcal{T}}_m^w ,
    \end{equation}
    where $E_0$ is the initial energy level of the battery, and $\eta_{\text{avg}}$ is the average charging/discharging efficiency. The final energy level ($E_{\text{des}}$), as set by the user, should be met with $\epsilon$ tolerance by time $T$ as
    \begin{equation} \label{e_final}
        \left|E_{\text{des}} - \left( E_0 + \eta_{\text{avg}}\Delta t\left(\sum_{j=1}^{T} P_{\text{bat},j} \right)  \right) \right| \leq \epsilon.
    \end{equation}
Note that constraints (\ref{e_bounds}) and (\ref{e_final})  couple the strategies of players V2G and BD, while constraint (\ref{p_bounds}) is local.
For compactness, we collect all the constraints in the set $\Omega_i$ given by:
\begin{equation}\label{feasibility}
    \Omega_1(u^{-1}):= \{u^1 \in \mathbb{R}^{m_1} \mid  \eqref{e_bounds} \ \text{and} \  \eqref{e_final} \  \text{hold}, \ \forall t \in \mathcal{T}_m^w\}, \nonumber 
\end{equation}
\begin{equation}\label{feasibility}
    \Omega_2(u^{-2}):= \{u^2 \in \mathbb{R}^{m_2} \mid  \eqref{e_bounds} \ \text{and} \  \eqref{e_final} \  \text{hold}, \ \forall t \in \tilde{\mathcal{T}}_m^w\}, \nonumber 
\end{equation}
and define the local feasibility sets $ U^1:= \{u^1  :  \eqref{p_bounds} \  \text{holds}, \ \forall \ t \in \mathcal{T}_m^w\} $ and $ U^2:= \{u^2  :  \eqref{p_bounds} \  \text{holds}, \ \forall \ t  \in \tilde{\mathcal{T}}_m^w\}$.

Player V2G then solves the optimization problem 
\begin{align} \label{obj_1}
  \min\limits_{u^1 \in U^1} \quad &  \theta^1 (u^1,u^{2}) := \sum\limits_{i\in [T]} \alpha_i(P_{\text{bat},i}\Delta t) \nonumber \\
  \text{ s.t. } \quad & u_1 \in \Omega_1(u^2),
\end{align}
where $\alpha_t$ (in \euro/kWhr) is the V2G price\footnote{This dynamic price inherently encodes different ancillary services offered to the grid, which include demand response, peak shaving, and congestion management.} as provided by the charging operator, and $\Delta t$ is the fixed time interval. 
Player BD solves the optimization problem
\begin{align}\label{obj_2}
          \min\limits_{u^2 \in U^2} \quad &\theta^2 (u^1,u^{2}) := \gamma \left(\sum\limits_{i\in [T]}  \: Q_{\text{loss},i}^{\text{cyc}}(P_{\text{bat},i})\right) \nonumber \\
          \text{ s.t. } \quad & u_2 \in \Omega_2(u^1),   
\end{align}
where $\gamma$ is the weight associated to battery health degradation, and $Q_{\text{loss},i}^{\text{cyc}}$ is the capacity loss at the $i^{\text{th}}$ interval as given in \eqref{loss_model}. Note that, $\theta^2$ does not take into account the calendar loss $Q^{\text{cal}}_{\text{loss}}$ from equation \eqref{calendar_eq}. This modelling choice stems from the fact that $Q^{\text{cal}}_{\text{loss}}$ remains constant for given values of $T_b$ and $t$. The calendar loss is thus present, irrespective of the EV usage.  As such, we focus solely on the cyclic battery aging $Q^{\text{cyc}}_{\text{loss}}$, since, contrary to the calendar aging,  $Q^{\text{cyc}}_{\text{loss}}$  directly depends on the power rate $P_{\text{bat},t}$, which constitutes a decision variable in our methodology. We denote the game between players V2G and BD by $\bar{\mathscr{G}}$.

\begin{figure}[h]
    \centering
    \includegraphics[width=0.8\linewidth]{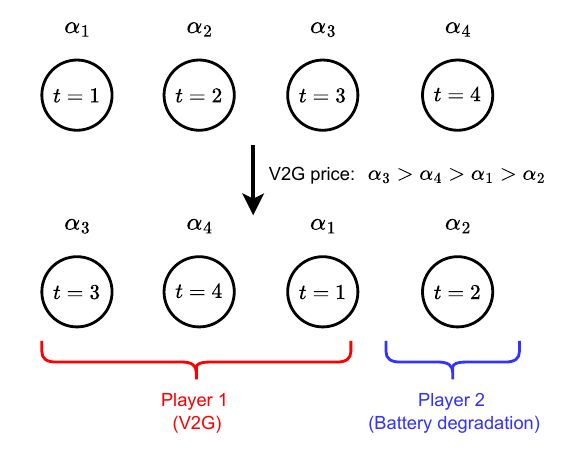}
    \caption{ \small Assuming $T=4$ and $w=3$, the allocation of time intervals to both players is shown. Player V2G gets $3$ intervals of relatively higher V2G price $(\alpha_t)$, and the remaining interval goes to player BD.}  
    \label{horizon_splitting}
\end{figure}

\begin{remark}
    \textcolor{black}{Previous studies \cite{zheng2015effects,rahman2024exploring} have demonstrated that battery capacity fade is influenced by the average state of charge (or equivalently, state of energy, $E$) and the depth of discharge (DoD). However, the effects of average state of charge and DoD on battery degradation are highly dependent on individual electric vehicle (EV) usage patterns. In particular, we note that: (a) parameters such as the initial energy level ($E_0$) and desired energy level ($E_{\text{des}}$) are user-defined and inherently stochastic; and (b) variations in driving behavior significantly influence how average state of charge and DoD affect battery health. As a result, our vehicle-to-grid (V2G) framework has limited control over these factors. Therefore, we do not explicitly incorporate average state of charge or DoD into our battery degradation model.}
\end{remark}
In our game-theoretic formulation, the hyperparameters $w$ and $m$ provide two degrees of freedom when choosing a solution. By varying $w$ keeping $m$ fixed, we essentially prefer one objective over another, and by varying $m$ keeping $w$ fixed, we change the quality of the solution. We can choose $m$ in $\Comb{T}{w}$ combinatorial ways. Thus, our horizon-splitting model provides a more flexible setting when it comes to choosing an appropriate smart charging strategy compared to other approaches. We propose the following price-based interval assignment strategy for choosing one of the $m$ combinations. 
 \begin{enumerate}
 \item Once $w$ is chosen by the EV user, sort the $T$ intervals based on the value of $\alpha_t$ in descending order. 
 \item Assign the first $w$ intervals to player V2G and the rest $(T-w)$ intervals to player BD, keeping the order of sorting intact.
 \end{enumerate}
A schematic diagram is given in Fig. \ref{horizon_splitting} to better illustrate the strategy. The main idea behind devising the above strategy is to allocate the intervals with a higher price ($\alpha_t$) to player V2G so that it can exploit the higher price value to its advantage. However, note that if $(T-w)$ is small, then the impact of choosing an optimal $m$ becomes minimal. 

Thanks to the convexity and compactness of the problem data, there always exists a GNE solution for our model. Furthermore, we show next that our game $\bar{\mathscr{G}}$ in \eqref{obj_1}, \eqref{obj_2} is an exact potential game. 
\begin{comment}
\begin{align*}
    (F(x)-F(y))^{\top} (x-y) > 0, \quad \forall x,y \in \Omega, \text{and} \: x\neq y,
\end{align*}
\end{comment}
 
\begin{proposition}\label{prop} (Potential game characterization)
    The GNE problem $\bar{\mathscr{G}}$ in \eqref{obj_1}, \eqref{obj_2}, is an exact generalized potential game with potential function 
    \begin{align}
    P(\mathbf{u}):= \sum_{j\in \mathcal{T}_m^w} (\alpha_j P_{\text{bat},j} \Delta t) + \gamma\sum_{j \in \Tilde{\mathcal{T}}_m^w} Q^{\text{cyc}}_{\text{loss},j}(P_{\text{bat},j}).
    \end{align}
%     \hfill $\square$ \par 
\end{proposition}
\emph{Proof:} See Appendix B.  \par 
    
Based on Proposition \ref{prop}, the following optimization problem provides a Nash equilibrium solution to the game $\bar{\mathscr{G}}$ in \eqref{obj_1}, \eqref{obj_2}. 
\begin{align}\label{gpg}
 \mathscr{G}_P:   \min_{\mathbf{u} \in U \times \Omega} \quad & P(\mathbf{u}) 
\end{align}
where $\mathbf{u}:=\text{col}((P_{\text{bat},t})_{t \in [T]})$,
$\Omega:= \{\mathbf{u}:  \eqref{e_bounds} \ \text{and} \  \eqref{e_final} \  \text{hold}, \ \forall t \in [T]\} $ and 
$U:= \{\mathbf{u}  :  \eqref{p_bounds} \  \text{holds}, \ \forall t \in [T]\}$.
Given that the problem in \eqref{gpg} is a convex optimization problem, we can retrieve a globally optimal solution, which corresponds to a GNE of $\bar{\mathscr{G}}$. In what follows, we propose a different formulation of the decision problem at hand based on multi-objective optimization.
\begin{remark}
    In general, the choice of the hyperparameter $w$ is based on the subjective importance users assign to the battery degradation. However, as a rule of thumb, one could select the $w^*$ that corresponds to the minimum of the sum of battery degradation and charging cost. Consider $u^*_w$ as the solution of \eqref{gpg} parameterized in $w$.  Therefore, $w^* := \argmin_{w \in [T]} \: [\theta^1(u^*_w)+ \theta^2(u^*_w)]$ could well be used as a reference guide for the user.  
\end{remark}
\subsection{Robustness of V2G charging: game theory vs multi-objective optimization}
We model a trade-off between the two conflicting objectives of V2G profit maximization and battery degradation minimization using a multi-objective optimization approach. In this case, a user-defined weight is assigned to each objective, reflecting their relative importance to the user. We frame our problem in terms of a multi-objective optimization problem as follows. 
\begin{align} 
    \label{mo}
    \min_{\mathbf{u} \in U \times \Omega} \quad  J(\mathbf{u}):= \: \left[ \rho \theta^1 (u^1,u^2) + (1-\rho) \theta^2 (u^1,u^2)\right] 
\end{align}
where $u$, $\Omega$ and $U$ are defined as in the formulation of $\mathscr{G}_P$ in (\ref{gpg}). $\rho \in [0,1]$ is the weight chosen by the user to indicate which objective among V2G revenue and battery degradation health is more favorable. We want to investigate the outcome of the game-theoretic approach (Section III.B) and the multi-objective optimization approach. To this end, we evaluate both methodologies based on the same case study. Note that, for the capacity loss model of the EV battery in \eqref{loss_model}, the parameters $B_{1,t}$, and $B_{2,t}$ change after several cycles of operation. Thus, a sensitivity study of both solution approaches with respect to $B_{1,t}$ and $B_{2,t}$ is of utmost importance.    
To study the robustness properties of both approaches, we first solve \eqref{gpg} and \eqref{mo} and retrieve their solutions $x_{\text{gt}}$ and $x_{\text{mo}}$ respectively, i.e.,  
\begin{subequations} \label{optimizer}
    \begin{align}
        & u_{\text{gt}}(w,\zeta)= \argmin_{\mathbf{u} \in U \times \Omega} \: \tilde{P} (\mathbf{u},\zeta), \\
        & u_{\text{mo}}(\rho,\zeta)= \argmin_{\mathbf{u} \in U \times \Omega} \: \tilde{J}(\mathbf{u},\zeta),
    \end{align}
\end{subequations}
where  $\tilde{P}(\cdot)$, and $\tilde{J}(\cdot)$ are the perturbed versions of $P(\cdot)$ and $J(\cdot)$ respectively; $\rho \in [0,1]$, $w\in[T]$, and $\zeta \in [\underline{a}, \overline{a}]$ is an uncertain perturbation parameter. Given $\{u_{\text{gt}}, u_{\text{mo}}\}$, we define the sensitivity functions as follows:     
\begin{subequations}
    \label{sensitivity_eq}
    \begin{align} 
        S_{\text{gt}} (w,\zeta) & := \frac{\left \Vert u_{\text{gt}}(w,\zeta_0) - u_{\text{gt}}(w,\zeta)\right \Vert}{\left \Vert\zeta - \zeta_0\right \Vert}, \\
        S_{\text{mo}} (\rho,\zeta) & := \frac{\left \Vert u_{\text{mo}}(\rho,\zeta_0) - u_{\text{mo}}(\rho,\zeta)\right \Vert}{\left \Vert\zeta - \zeta_0\right \Vert}  ,
    \end{align}
\end{subequations}
where $\zeta_0$ is the nominal value of the uncertain parameter, $\left \Vert \cdot \right \Vert$ is the euclidian norm, and $\zeta$ follows an uniform distribution constructed on $[\underline{a}, \overline{a}]$. Such a sensitivity analysis is important because the trade-off insights derived from our study should remain meaningful to a certain extent, irrespective of the changes in battery condition. When we solve \eqref{gpg} or \eqref{mo} considering $\zeta_0$, we do not know the true realization of $\zeta$ in advance. This motivates us to investigate the impact of a different realization of $\zeta$ other than $\zeta_0$. We define the notion of empirical regret in the following sense: 
\begin{subequations}\label{reg_eq}
    \begin{align} \label{reg_gt}
        R_{\text{gt}} (w,\zeta) & := \frac{\tilde{P}(u_{\text{gt}}(w,\zeta_0),\zeta) - \tilde{P}(u_{\text{gt}}(w,\zeta), \zeta) }{\tilde{P}(u_{\text{gt}}(w,\zeta), \zeta)} \\
        R_{\text{mo}} (\rho,\zeta) & := \frac{\tilde{J}(u_{\text{mo}}(\rho,\zeta_0),\zeta) - \tilde{J}(u_{\text{mo}}(\rho,\zeta), \zeta)}{\tilde{J}(u_{\text{mo}}(\rho,\zeta), \zeta)}
    \end{align}
\end{subequations}
where the term $\tilde{P}(u_{\text{gt}}(w,\zeta), \zeta)$ (resp. $\tilde{J}(u_{\text{mo}}(\rho,\zeta), \zeta)$ ) signifies the objective function value corresponding to the solution when the decision maker knows the true realization of $\zeta$ in advance, and $\tilde{P}(u_{\text{gt}}(w,\zeta_0), \zeta)$ (resp. $\tilde{J}(u_{\text{mo}}(\rho,\zeta_0), \zeta)$ is the value of the objective function based on the solution where the decision maker considered $\zeta_0$. In other words, the empirical regret indicates the relative loss of the objective function value if changes in battery degradation parameters are not considered in the formulation. Thus, the solution approach with a smaller sensitivity or regret value is considered more reliable, especially when the horizon is long.    

\section{Numerical Simulations}
\subsection{Simulation Setup}
Since both optimization problems in \eqref{gpg} and \eqref{mo} are convex, we used the Python-based package Pyomo \cite{pyomo} to formulate the problem and solved them using the MOSEK \cite{mosek} solver. We consider the V2G tariff profile $\alpha_t$ of Fig. $7$ in \cite{lai2022pricing}, thus leveraging a dynamic pricing strategy for EVs to provide demand response-based ancillary services to the local distribution grid. Furthermore, we assume the EV user takes part in the V2G program between $8$ AM-$8$ PM. The discretizing interval ($\Delta t$) is $15$ minutes. Considering $T=48$ intervals (i.e., $12$ hours), MOSEK solver takes $ <50 \: \text{ms}$ to solve \eqref{gpg}. The processor used is \textit{Apple M1 Pro} with a 10-core CPU. Such a low computation time is due to the convexity of the optimization problems, which allows our proposed method to be implemented in real-time in an embedded platform with reasonable computational power. The upper ($\overline{E}$) and lower ($\underline{E}$) bounds of available energy in the battery are $1$ and $0.2$ (in p.u.) respectively. Similarly, maximum ($\overline{P}$) charging/discharging power is $22$ kW. The user tolerance $\epsilon$ is $0.02$ (in p.u.). The desired energy level of the battery ($E_{\text{des}}$) is $0.9$ (in p.u.). The capacity of a single cell battery $C_{\text{rated}}$ is $1.5$ Ah. Different parameters for $50$ kWh battery pack are $n_{\text{series}}$, $n_{\text{parallel}}$, and $n_{\text{max}}$ with the values of $83$, $94$, and $5.28$ respectively. The terminal voltage ($V_{\text{bat}}$) of the battery pack is $ 350$V.     
\begin{figure}
	\centering
	\begin{subfigure}{\linewidth}
    		\centering
    		\includegraphics[width=1.05\linewidth]{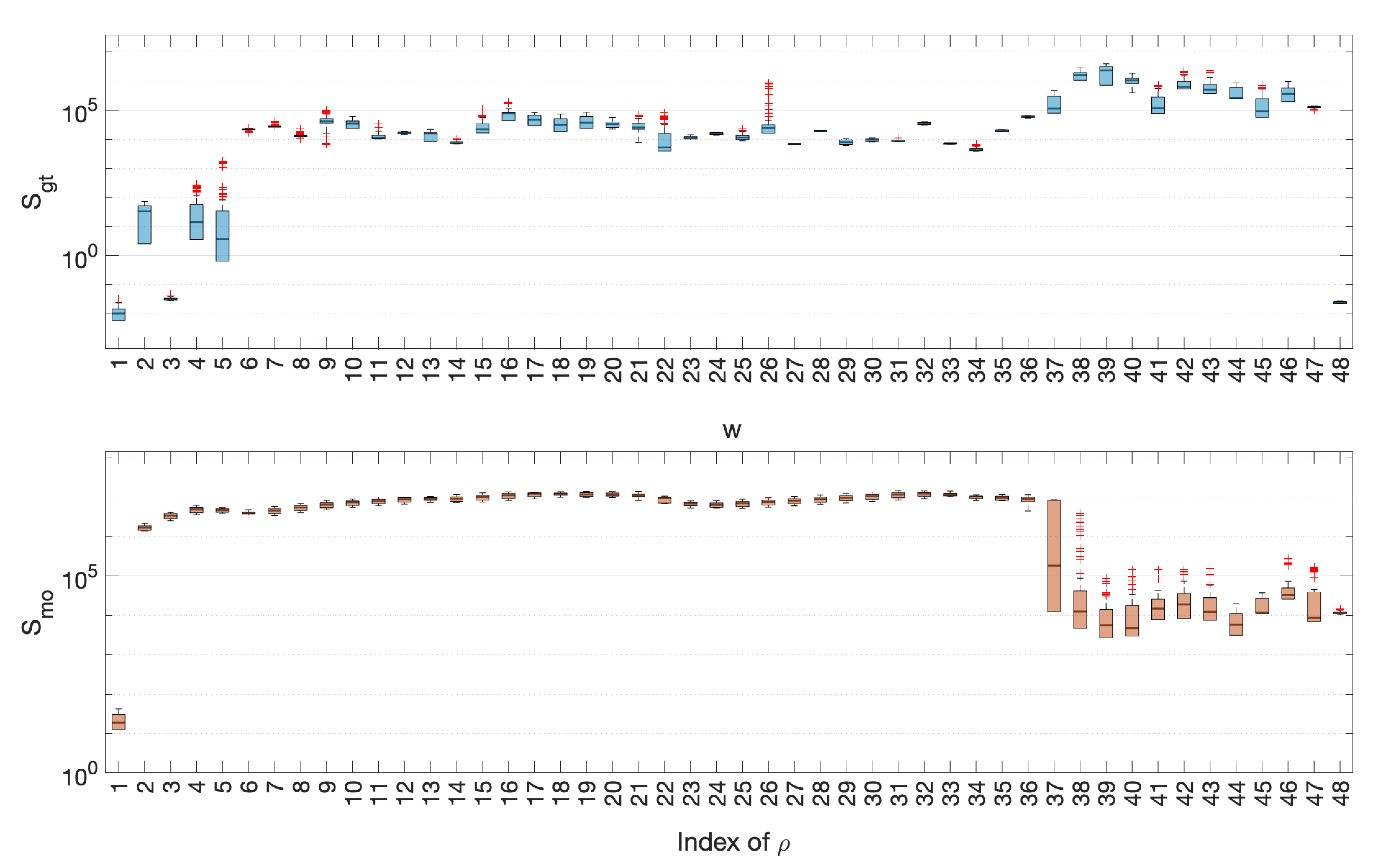}
                \caption{}
    		\label{sensi}
	\end{subfigure}
 
 	\begin{subfigure}{\linewidth}
    		\centering
    		\includegraphics[width=1.05\linewidth]{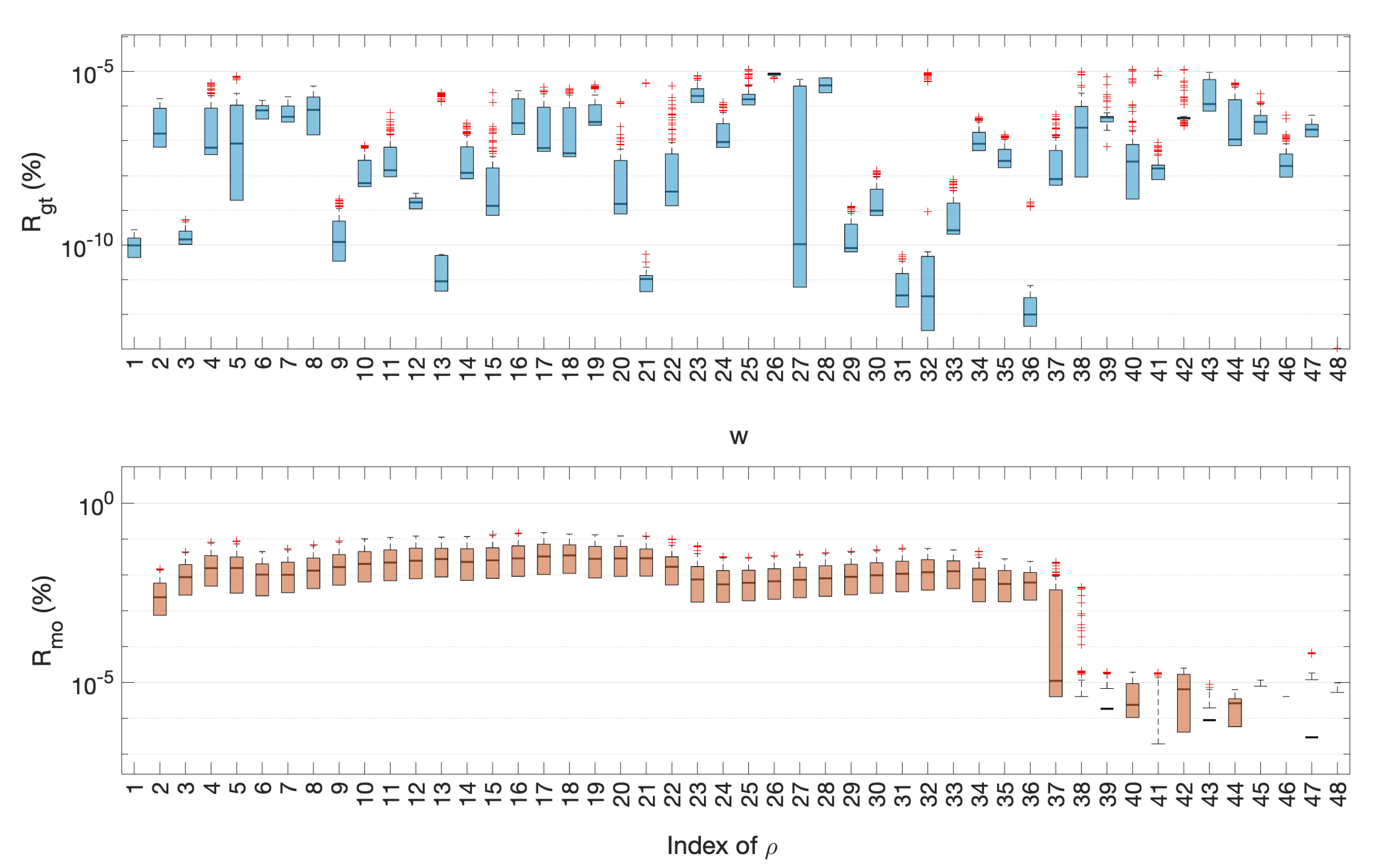} 
            \caption{}
            \label{regret}
	\end{subfigure}
	\caption{ \small (a) Box plots on the sensitivity of the solution profiles for the game-theoretic and multi-objective approaches, (b) box plot for regret evaluation of both approaches. }
\end{figure}

Given various standardized charging technologies available in the market, we are particularly focused on bi-directional EV chargers with variable charging/discharging rates as a recent study \cite{brinkel2024enhancing} demonstrates the financial benefits of it compared to charging with a fixed C-rate. For this purpose, we choose a level-2 three-phase on-board charger of power rating $22$ kW, which Eaton and other major manufacturers currently manufacture \cite{yuan2021review}. The market analysis in \cite{mohammadi2023comprehensive} shows that most passenger EVs have battery capacity in the range of ($50-100$) kWh. To account for the vast majority of users, we conduct the studies based on the battery capacity of $50$ kWhr in this subsection. Detailed results on how different capacity values affect the obtained results can be found in Subsection IV.C. For our simulation studies, we consider that a new battery costs $207$\euro/kWh (battery pack cost of Chevrolet Bolt in \cite{wentker2019bottom}). After a $30 \%$ loss in capacity, it can be further used as a second-life battery in other storage applications. Hence, we consider the resale value of it as $45$\euro/kWhr \cite{montes2022procedure}. The effective cost of our EV battery stands at $\frac{(207-0.7 \times45)}{0.3}=585$\euro/kWhr, which is the value of $\gamma$ in \eqref{obj_2}.  In the following, we construct several case studies to investigate the trade-off between battery degradation and the V2G exploitability of EVs.    
\subsection{ Comparative Study of Sensitivity \& Regret}
  
We use the notions defined in Section III.C to compare the empirical robustness of the game-theoretic and the multi-objective approaches by evaluating \eqref{sensitivity_eq} and \eqref{reg_eq}. In Fig. \ref{sensi}, the box plot of $S_{gt}$ and $S_{mo}$ is illustrated, for varying values of the hyperparameters $w$ (or, $\rho$), respectively\footnote{Because of the discrete nature of $w$, we considered $T$ number of discrete realizations of $\rho \in [0,1]$.}, and different realizations of the uncertain parameter $\zeta$. The parameter $\zeta$ is assumed to take values in the support set $[0.9\zeta_{0}, 1.1 \zeta_{0}]^n$ with $\zeta_{0}$ being the nominal value. The V2G session is set for twelve hours ($T=48$ intervals).  

Figure \ref{sensi} shows that the game-theoretic solution is significantly more robust compared to the multi-objective one. As a result, the trade-off insights derived from the game-theoretic solutions remain meaningful irrespective of the realizations of $\zeta$ in the support set compared to the multi-objective one. Finally, Fig. \ref{regret} depicts a comparison of the regret of the cumulative cost functions of both approaches as defined in \eqref{reg_eq}. We observe that, under perturbations in $\zeta$, the loss in the objective function value of the game-theoretic approach is significantly smaller compared to the multi-objective one. Having such low regret explains why our proposed method may not require further robustification against uncertainty, as opposed to a multiobjective approach. Thus, our approach avoids overly conservative decision-making without sacrificing safety against small perturbations. 

\subsection{V2G Exploitability and Battery Degradation Trade-off}
The trade-off is influenced by various factors, including average ambient temperature, EV battery capacity, charging duration, and the V2G tariff rate for energy transactions. However, we concentrate on three key factors that significantly affect both players.
\subsubsection{Impact of Ambient Temperature}
\begin{figure}[h]
    		\centering
    		\includegraphics[width=1\linewidth]{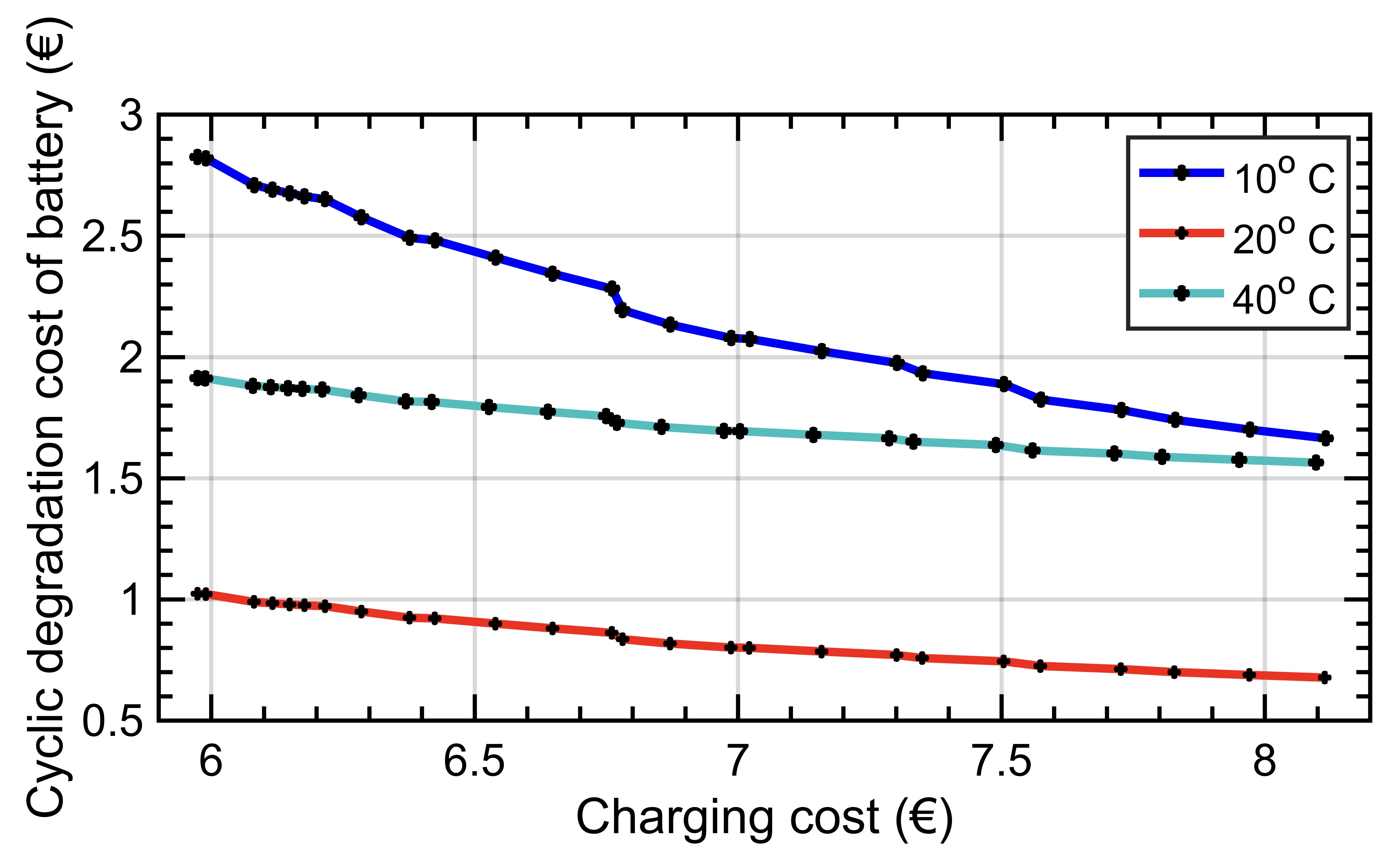}
                \caption{\small Trade-off curve for comparison between charging cost and cyclic battery degradation cost while taking part in V2G for different $T_a$. The markers along the curve indicate discrete solutions of \eqref{gpg} for decreasing values of $w$.   
                }
    		\label{temp_trade_off}
 \end{figure}

\begin{figure}[h]
    		\centering
    		\includegraphics[width=1\linewidth]{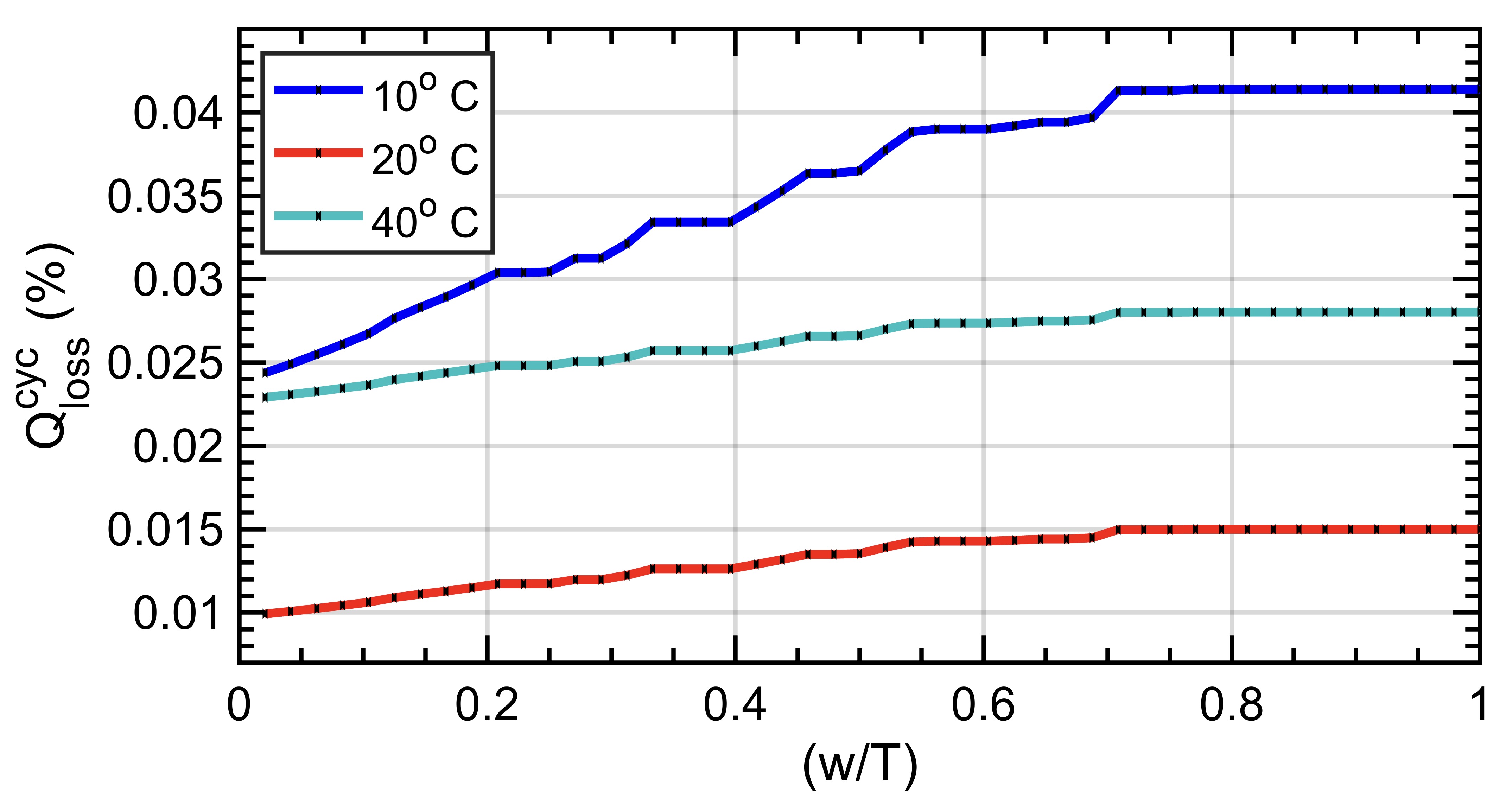}
	\caption{  \small The capacity loss for different choices of $w$ by the user under different $T_a$.   
    }
        \label{temp_curve}
\end{figure}

As discussed in Section II.C, battery temperature ($T_b$) plays a critical role in battery health degradation. EVs are operated globally across diverse climate zones, each with distinct average ambient temperatures ($T_a$) that vary by season. To account for this, we consider average $T_a$ values of $10^\circ$C, $20^\circ$C, and $40^\circ$C to represent EV usage in cold \footnote{In latest generation EVs, battery pre-heating is conducted by battery thermal management system \cite{hu2020battery} to bring back $T_b$ at $15^{\degree}$ C before operation in cold geographical locations. Our analysis does not capture this phenomenon. Instead, it is suitable for existing EVs devoid of such features.}, mild, and hot regions, respectively. The influence of $T_a$ is illustrated in Fig. \ref{temp_trade_off} and \ref{temp_curve}.
Note that in Fig.  \ref{temp_trade_off} the markers represent different user-defined trade-offs on the charging cost and battery degradation cost, controlled by the hyperparameter $w$. Specifically, the markers on the left-hand side of the plot correspond to the costs when the user has chosen a high value of $w$, i.e., they are more interested in minimizing the V2G charging cost without being concerned about the battery degradation. Respectively, the markers on the right-hand side of the figure correspond to lower values of $w$ and represent the case when the user prioritizes their vehicle's battery health over minimizing the V2G charging cost. 

We observe that for $T_a=20^\circ$C, putting more emphasis on minimizing the charging cost of V2G leads to more savings in the long run than being concerned with the vehicle's battery degradation. However, this pattern starts changing as the environmental temperature decreases. In fact, for colder regions with $T_a<10^\circ$C, EV users should exercise greater caution when selecting a higher value of $w$, i.e., placing more emphasis on V2G exploitability as the battery degradation costs can be significantly higher in the long run, as evidenced by the slope of the curve. Figure \ref{temp_curve} illustrates the percentage battery capacity loss due to cyclic aging as a function of the hyperparameter $w$ for different temperatures. A key observation is that as $T_a$ increases, the slope of $Q^{cyc}_{loss}$ with respect to $(w/T)$ decreases. This indicates that changes in V2G participation levels have a smaller impact on cyclic battery degradation in higher temperature zones.
\subsubsection{Impact of Varying V2G Tariff Profiles}
\begin{figure}
 
 	\begin{subfigure}{\linewidth}
    		\centering
    		\includegraphics[width=1\linewidth]{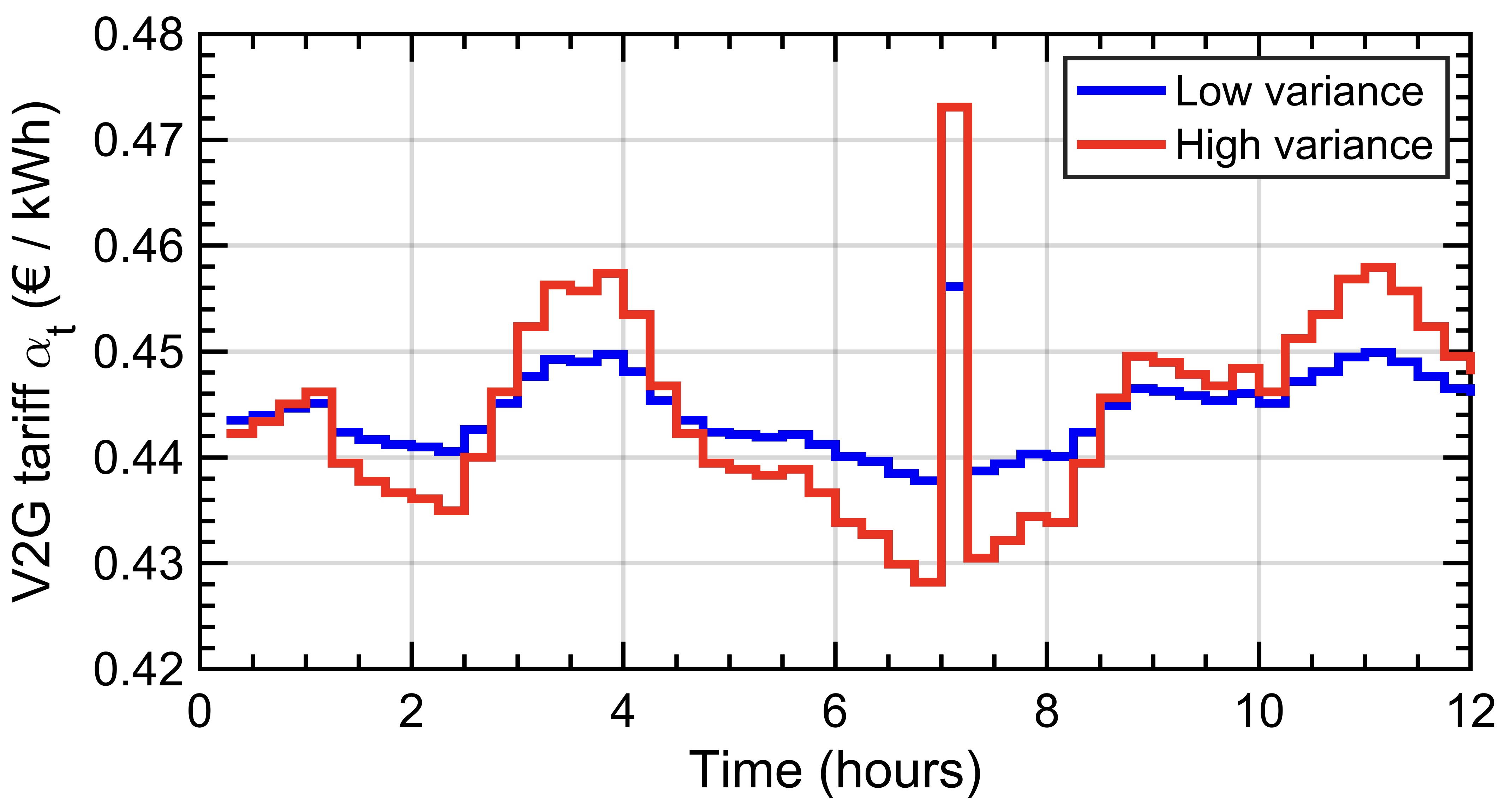} 
            \caption{}
            \label{variance}
	\end{subfigure}
        \begin{subfigure}{\linewidth}
    		\centering
    		\includegraphics[width=1\linewidth]{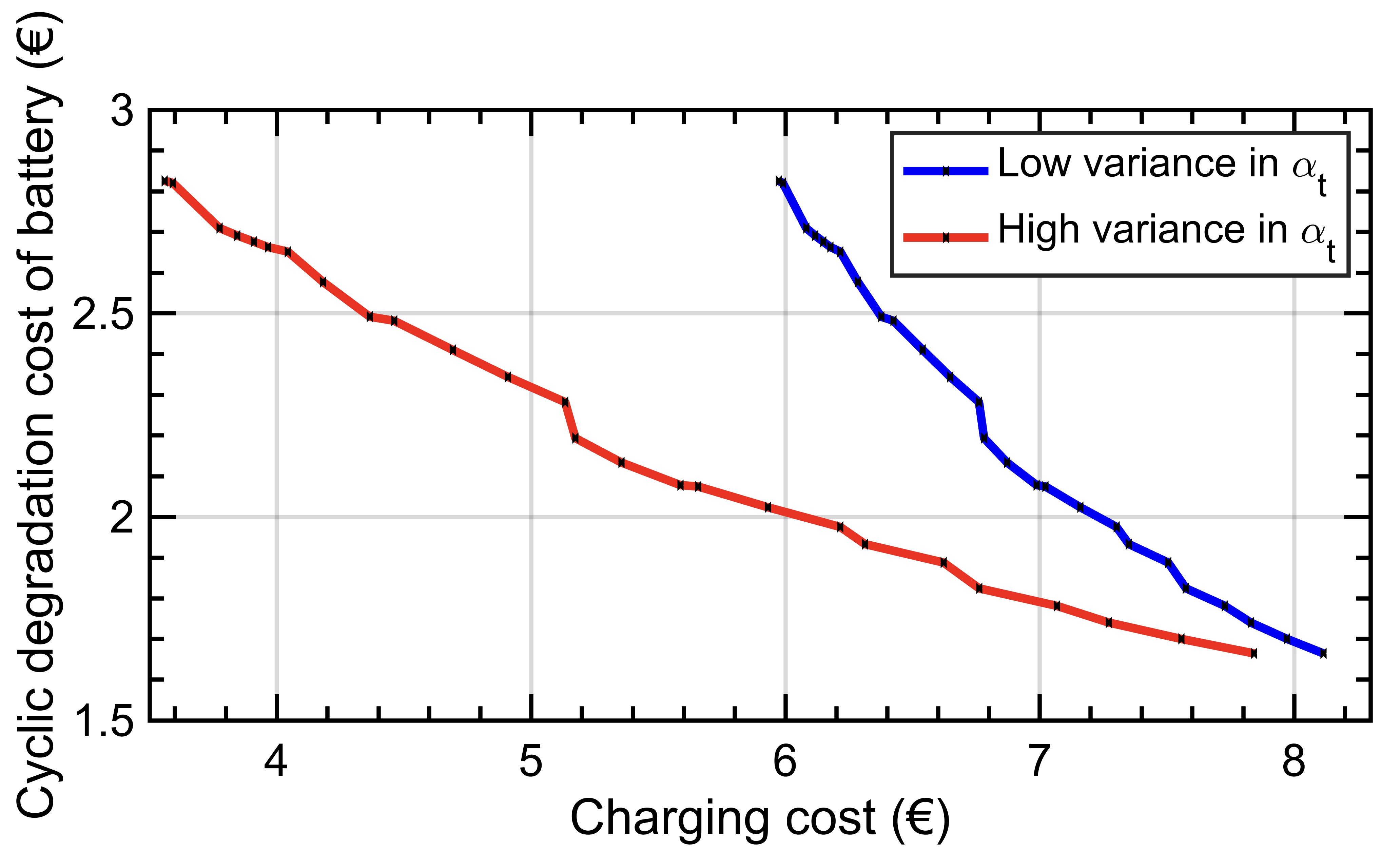} 
            \caption{}
            \label{trade_off_var}
	\end{subfigure}
	\caption{\small (a) Profiles of V2G tariff ($\alpha_t$) for different variances, (b) trade-off curve for comparison between charging cost and cyclic battery degradation cost while taking part in V2G for different variances in $\alpha_t$. }
        
\end{figure}

The V2G tariff profile $\alpha_t$ can exhibit different levels of variance depending on the flexibility available at the charging station or parking lot and the ancillary service requirements of local distribution system operators. A high variance profile characterized by frequent fluctuations and greater distances between consecutive peaks and troughs is depicted in Fig. \ref{variance}. Such profiles are common in regions with a high share of intermittent renewable energy sources and a large population of EVs. Conversely, regions without these conditions may experience a relatively low variance in the $\alpha_t$ profile. The level of variance in $\alpha_t$ significantly impacts the trade-off curve, as shown in Fig. \ref{trade_off_var}. Higher variance allows users to exploit V2G more effectively for monetary gains. As illustrated in Fig. \ref{trade_off_var}, users can opt for a higher value of $w$ while accounting for battery degradation costs under a high variance $\alpha_t$ profile. 
\subsubsection{Impact of Charger Power Ratings}
\begin{figure}
    \centering   \includegraphics[width=1\linewidth]{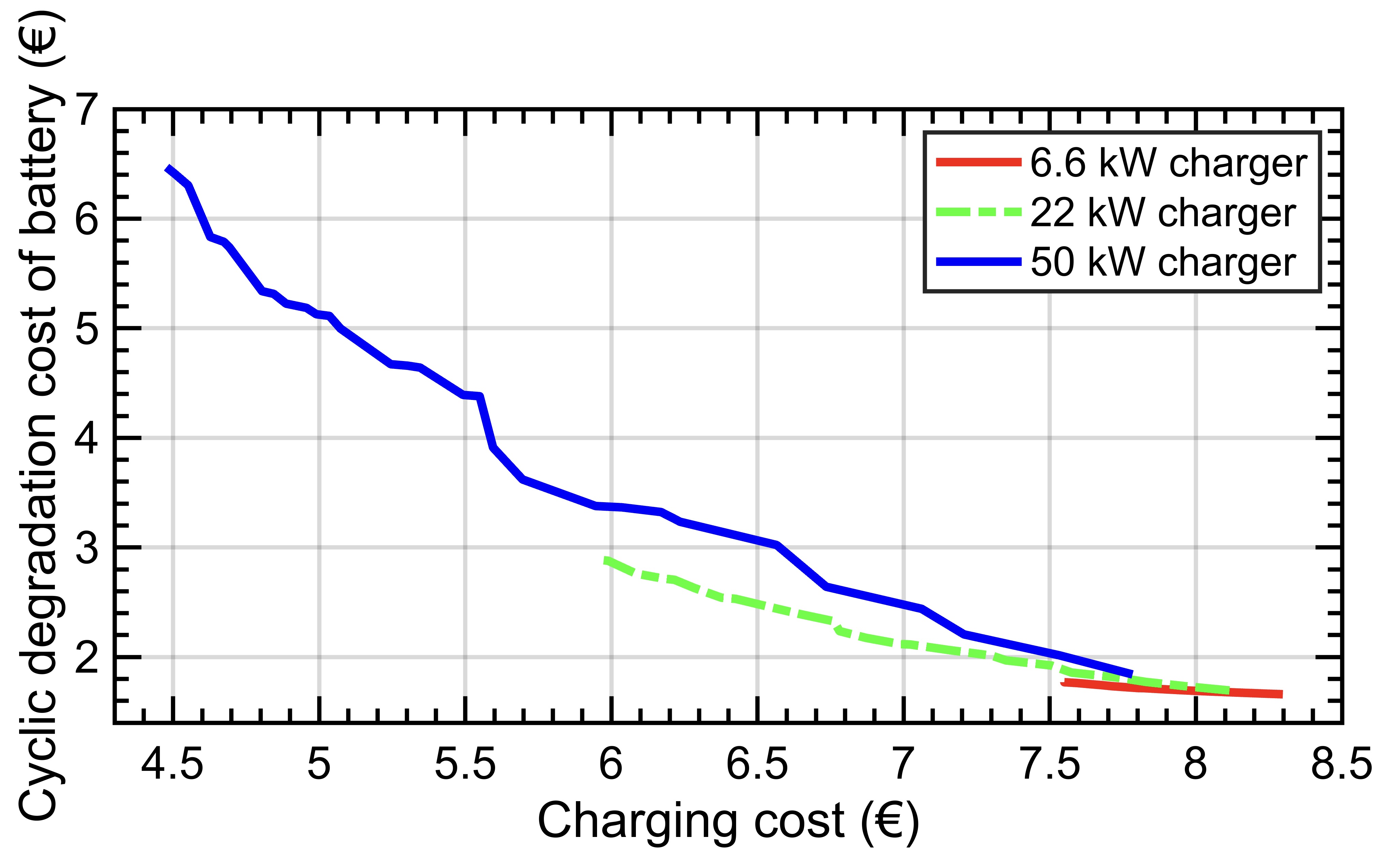}
    \caption{ \small Trade-off curve for comparison between charging cost and cyclic battery degradation cost for chargers of different power ratings.}
    \label{charger_plot}
\end{figure}

Currently, common EV chargers are classified into Level$-1$ (up to $7$ kW), Level$-2$ ($7-22$ kW), and Level$-3$ ($\geq50$ kW) categories. We investigate the trade-off when the same EV is charged using chargers with power ratings of $6.6$ kW, $22$ kW, and $50$ kW. The results, depicted in Fig. \ref{charger_plot}, show that variations in both charging cost and battery degradation cost are more pronounced when using higher-rated chargers. This indicates that both V2G exploitation and the extent of battery health degradation accelerate as the charger power rating increases. Therefore, higher-rated EV chargers offer users greater flexibility in selecting their preferred operating point on the trade-off curve.   
\subsection{Smart Charging Profiles of EVs}
\begin{figure}
    \centering \includegraphics[width=1\linewidth]{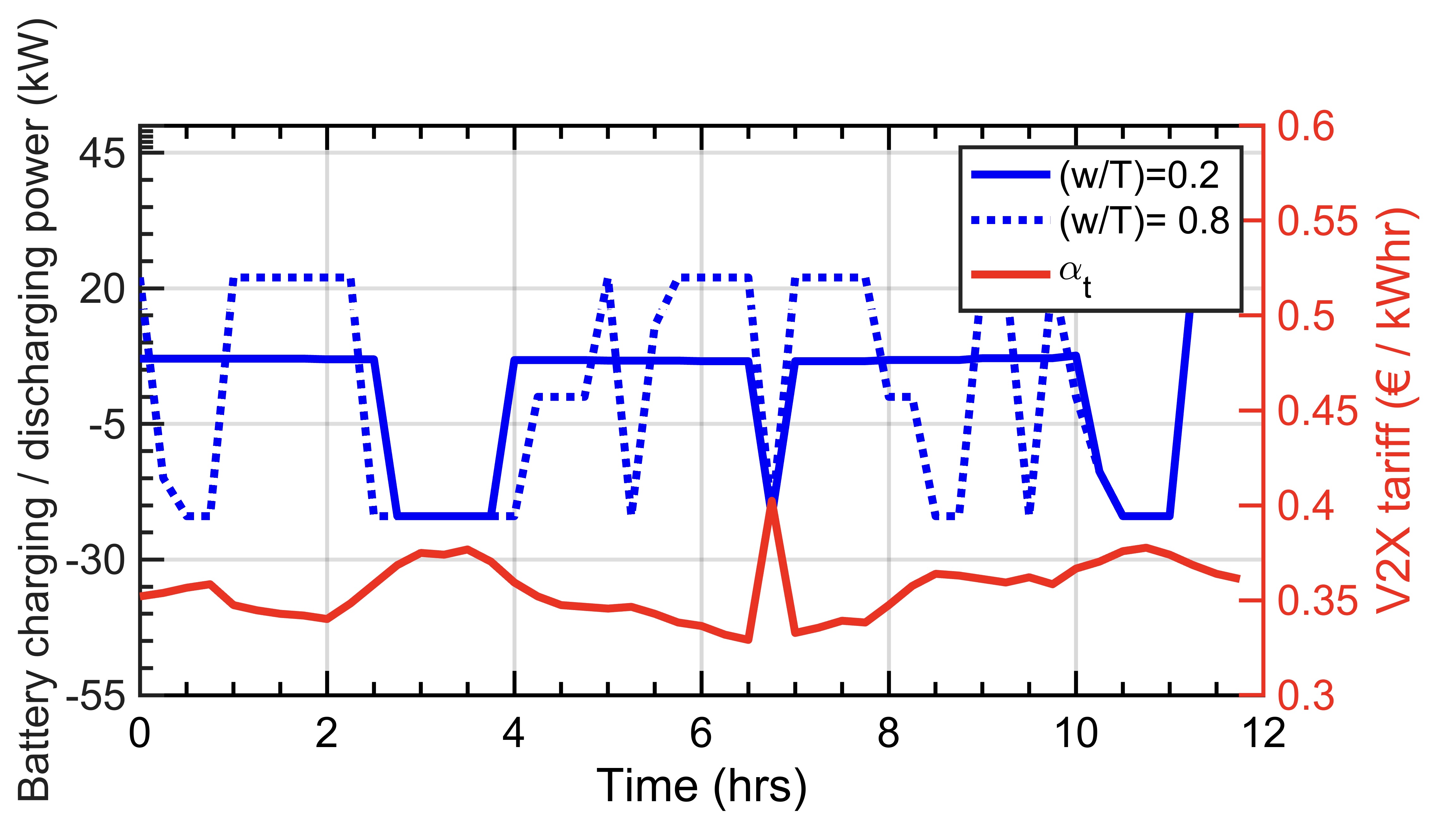}
    \caption{\small Smart charging profiles for different choices of $w$.}
    \label{charging_profile}
\end{figure}

Figures \ref{temp_trade_off} illustrate how users can select their desired operating point by adjusting the hyperparameter $w$ in our proposed framework. Figure \ref{charging_profile} depicts the charging power of an EV as $w$ varies, alongside the corresponding V2G tariff profile ($\alpha_t$) on the secondary y-axis. Player V2G aims to minimize total charging costs by purchasing energy at low prices and selling it back to the grid at higher prices. Conversely, player BD seeks to reduce the C-rate of charging and discharging, resulting in a relatively flat charging profile with minimal exploitation of the $\alpha_t$ profile. Notably, regardless of the $w$ values, the final energy level consistently approaches the user's desired energy level $E_{\text{des}}$.   
\subsection{Long-term Trade-off Analysis for Different Battery Capacities} 
\begin{figure}
	\centering
	\begin{subfigure}{\linewidth}
    		\centering
    		\includegraphics[width=1\linewidth]{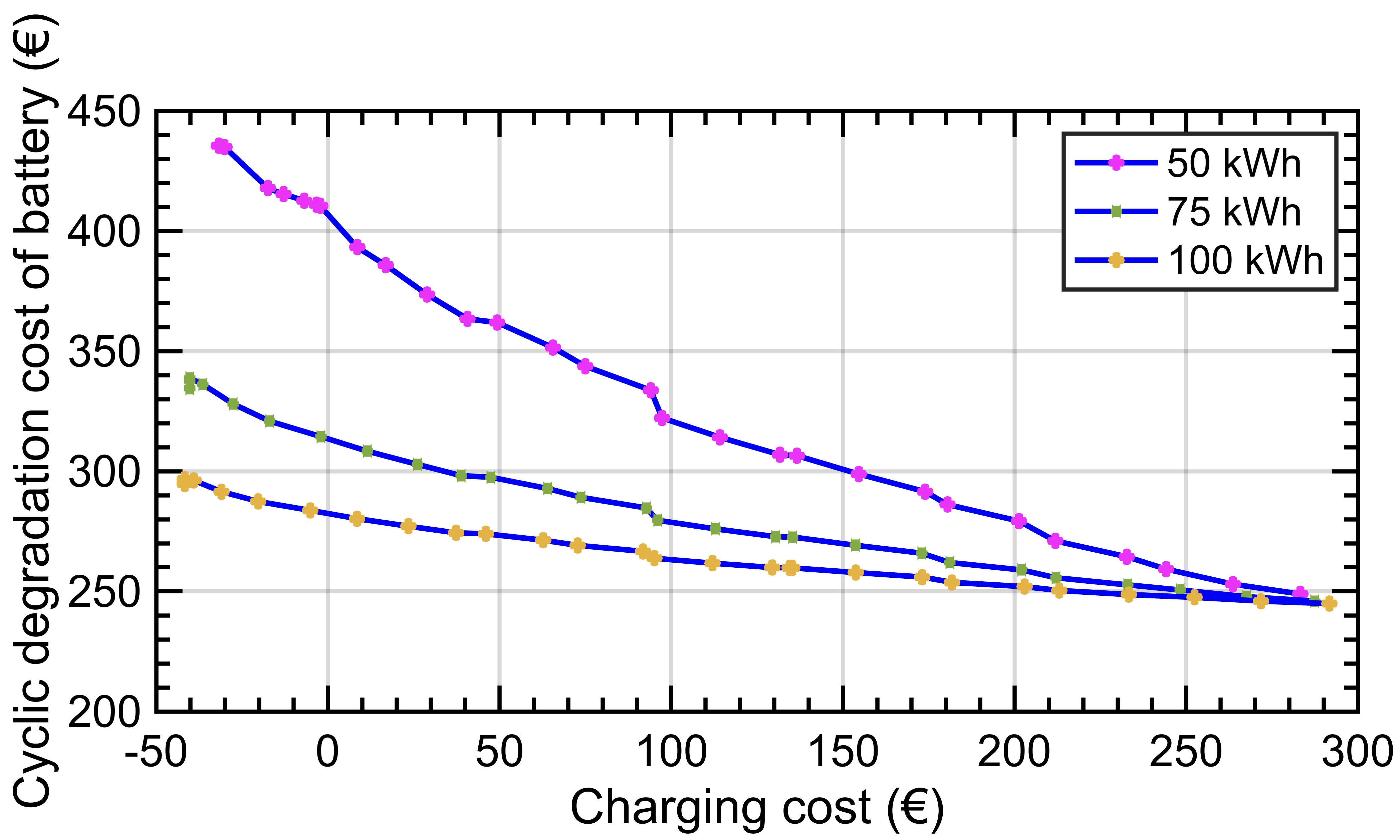}
                \caption{}
    		\label{proj_plot}
	\end{subfigure}
 
 	\begin{subfigure}{\linewidth}
    		\centering
    		\includegraphics[width=1\linewidth]{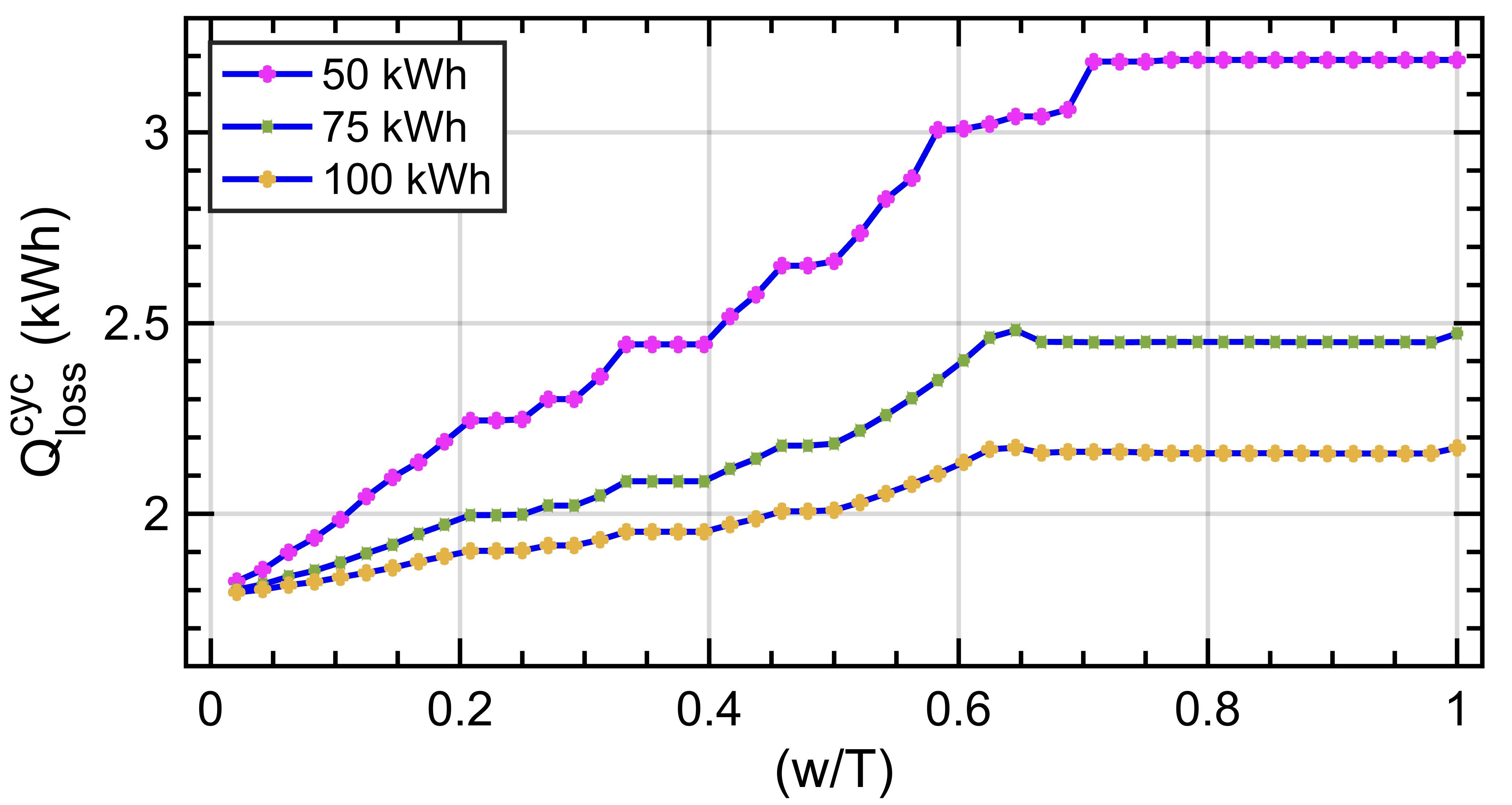} 
            \caption{}
            \label{proj_cap}
	\end{subfigure}
	\caption{ \small (a) Trade-off curve of the one-year-long projection study where the EV battery capacities are varied, and (b) the corresponding battery capacity losses as the user-defined hyperparameter $w$ varies.}
\end{figure}

A crucial aspect of this work involves studying the \emph{long-term} effect of V2G services on battery degradation, i.e., how a gradual, seemingly negligible accumulation of daily degradations due to V2G can affect battery health in the long run. To this end, we conducted a projection study using one year of real data for EVs with battery capacities of 50 kWh, 75 kWh, and 100 kWh. The setup for the year included the following:

The EV user drives an average of 30 km daily, resulting in approximately 5 kWh of energy consumption per day \cite{andersen2018added}. The EV user participates in a V2G program three days a week, with each session lasting 12 hours. One year’s worth of ambient temperature data ($T_a$) for Delft, the Netherlands, was used, as sourced from \cite{temp_data}. Each data point in the daily V2G tariff profile was sampled from a Gaussian distribution, with the mean taken from Fig. $7$ of \cite{lai2022pricing} and a standard deviation of 10$\%$. This setup was chosen to reflect a realistic scenario, allowing the trade-off analysis to yield practical insights. Numerical simulations based on this data produce the trade-off plot shown in Fig. \ref{proj_plot}, which reveals that the cyclic degradation cost becomes increasingly competitive with the charging cost as battery capacity decreases.\footnote{Note that a positive charging cost in Fig. \ref{proj_plot} indicates that the EV user incurs a cost, whereas a negative cost indicates that the EV user is compensated by the V2G service provider at the end of the year.} Note in both figures that the number of charging/discharging cycles increases as battery capacity decreases, leading to greater cyclic degradation by the end of the year. Secondly, the ratio of battery capacity to charger power rating plays a significant role in battery degradation. A lower ratio results in higher cyclic degradation because the charging/discharging cycles occur at a higher C-rate for smaller batteries when using the same charger. Previous studies \cite{vermeer2021comprehensive,wang2014degradation} have shown that cycling a battery with a higher C-rate causes increased battery health degradation. The key takeaway from Fig. \ref{proj_plot} is that EV users with lower-capacity batteries must carefully manage their level of V2G participation. For larger-capacity batteries, full participation in V2G is likely to result in minimal additional battery degradation costs relative to the monetary benefits gained.
\subsection{Comparison with MPC-based method in \cite{lu2024coordinated}}
\begin{figure}
    \centering \includegraphics[width=1.1\linewidth]{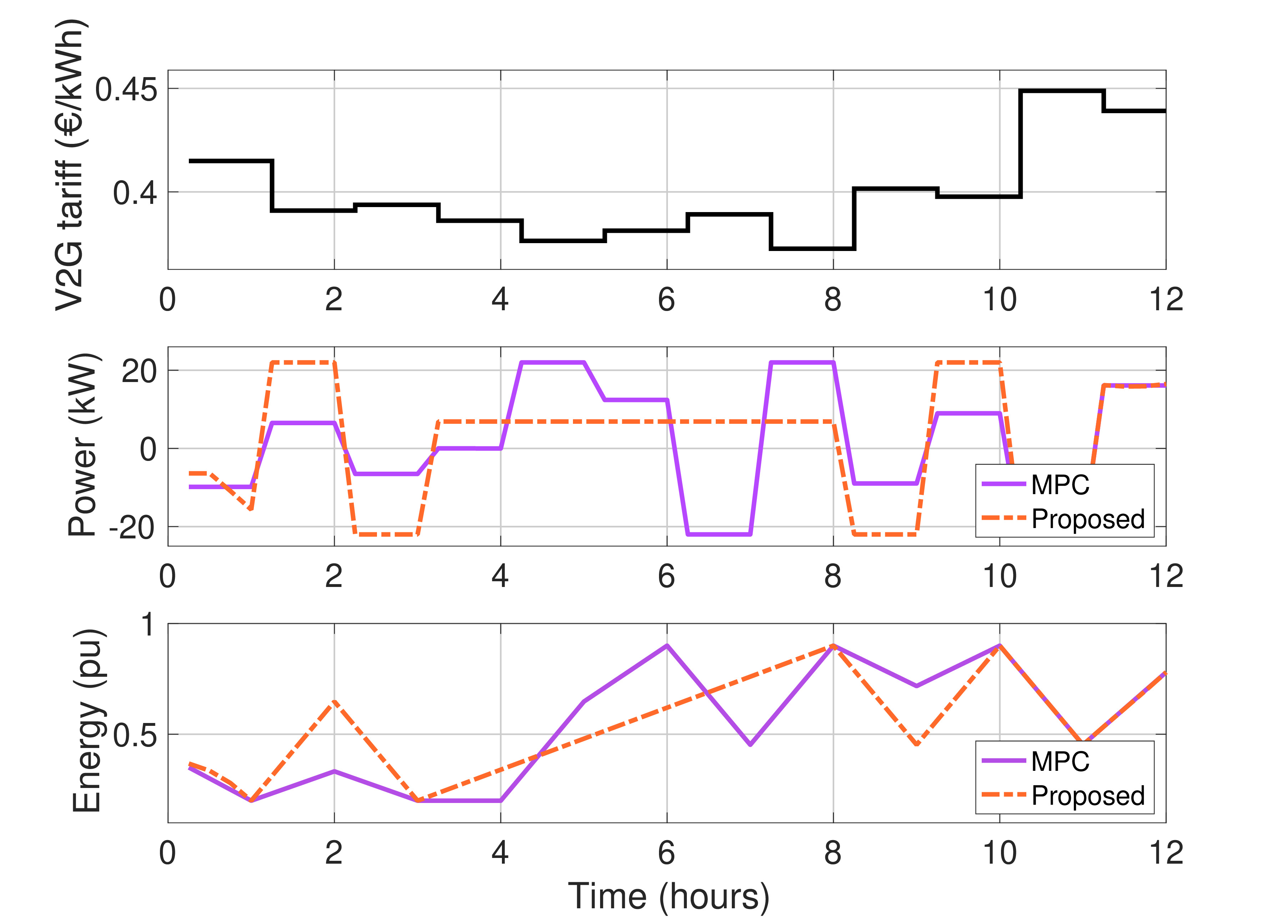}
    \caption{\small \textcolor{black}{Comparison of power and energy profiles based on:  1) MPC-based method \cite{lu2024coordinated} (solid violet) and 2) our proposed method (dashed orange). Both methods respect power constraints and reach the desired final energy $E_{\text{des}}$ of $0.8$, as set by the user.}}
    \label{compare_sol}
\end{figure}

In the recent work \cite{lu2024coordinated}, the authors propose a model predictive control (MPC)-based vehicle-to-grid (V2G) charging algorithm for electric vehicles (EVs), incorporating battery degradation into the optimization framework. Considering their algorithm is applied to each EV separately, it serves as a state-of-the-art benchmark to compare our method. \textcolor{black}{To ensure a fair comparison, we assume that the objective in \cite{lu2024coordinated} is to minimize the sum of the charging cost and the battery degradation cost. We also use the same battery degradation model for both methods. Furthermore, we use the same set of uncertain samples to evaluate various metrics.} In our game-theoretic framework, we assign equal importance to both cost components by setting $w=T/2$. Using identical V2G tariff profiles and constraint settings for both methods, the resulting solutions are presented in Fig. \ref{compare_sol}. As expected, both approaches satisfy all constraints and successfully achieve the user-specified terminal energy level $E_{\text{des}}$ by the end of the V2G session.

\begin{figure}[h]
	\centering
	\begin{subfigure}{\linewidth}
    		\centering
    		\includegraphics[width=0.8\linewidth]{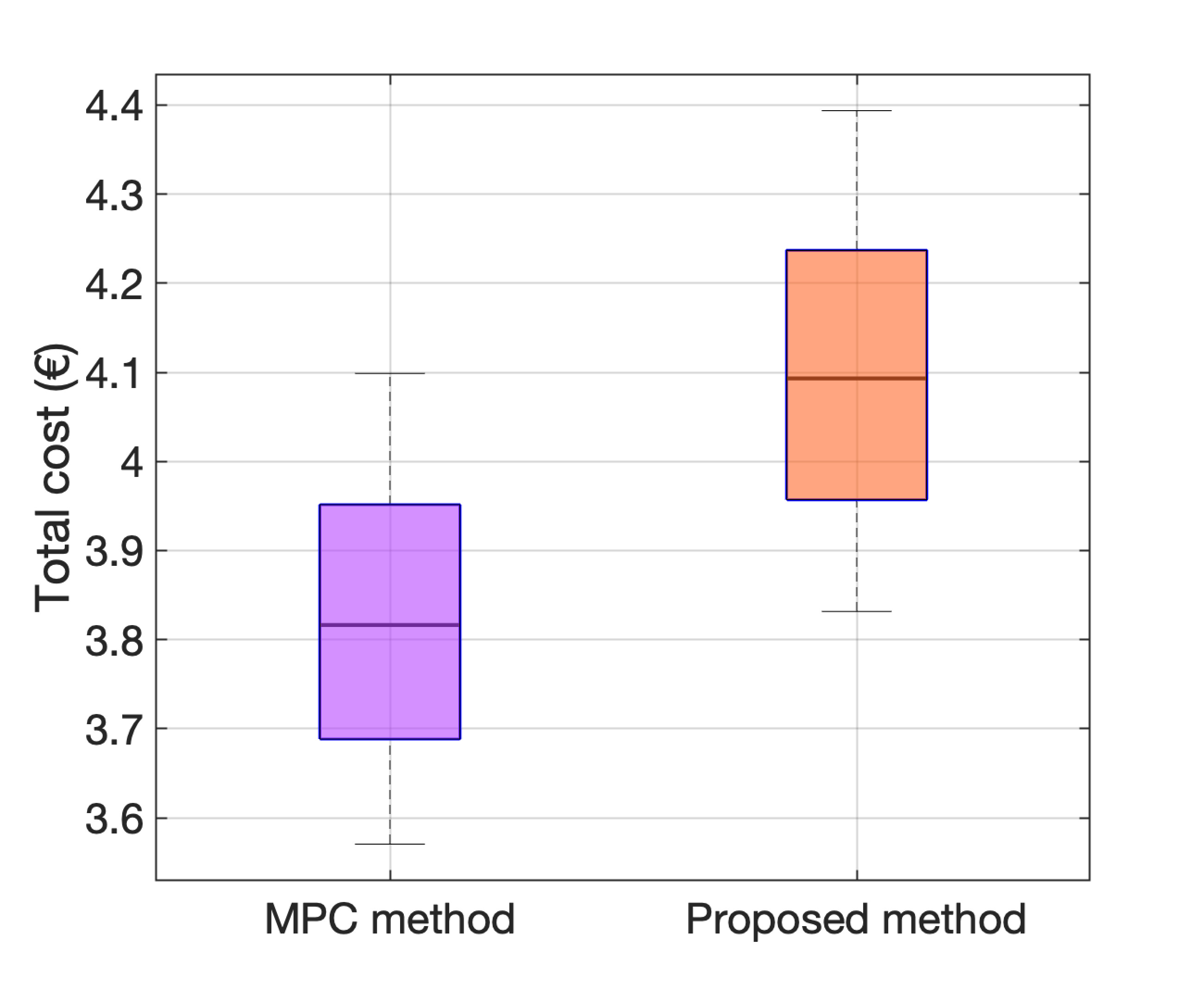}
                \caption{}
    		\label{comp_cost}
	\end{subfigure}
 
 	\begin{subfigure}{\linewidth}
    		\centering
    		\includegraphics[width=0.8\linewidth]{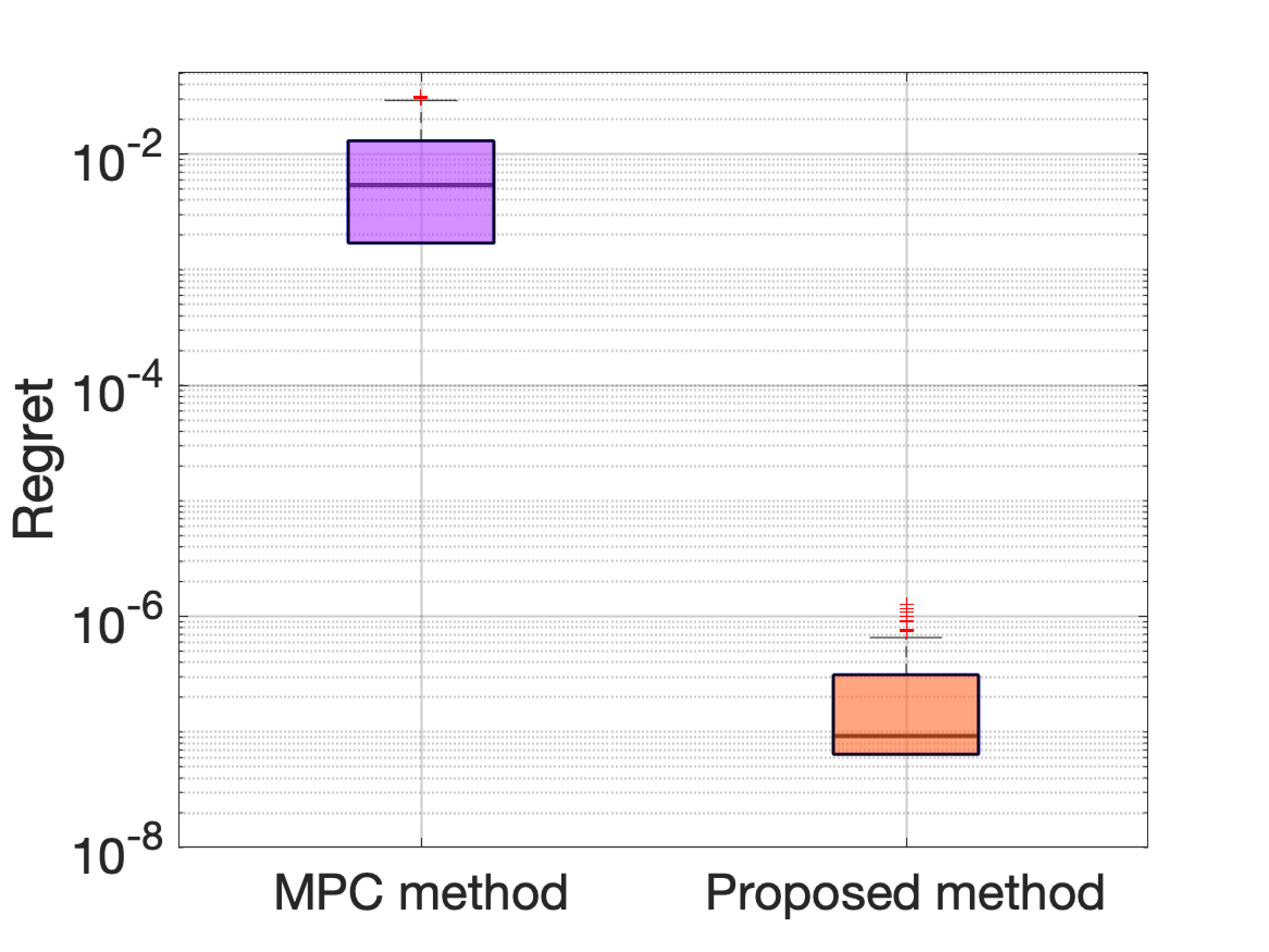} 
            \caption{}
            \label{comp_regret}
	\end{subfigure}
	\caption{ \small \textcolor{black}{ (a) Box plot for comparison of the total cost between MPC-based method \cite{lu2024coordinated} and the proposed method when $\zeta$ in $\theta^2(\cdot)$ is perturbed, (b) box plot in logarithmic scale for comparison of the regret metric (following \eqref{reg_eq}) between these methods. } }
\end{figure}

However, when we investigate the behaviour of total cost (charging cost $+$ battery degradation cost) as the uncertain parameter $\zeta$ in the battery degradation cost function ($\theta^2$) is perturbed, we observe the following: 1) Assuming $\zeta$ is sampled from a uniform distribution with support set $[0.9\zeta_{0}, 1.1 \zeta_{0}]^n$, Fig. \ref{comp_cost} shows that, our method incurs a slightly higher expected total cost compared to its counterpart. 2) On the contrary, when we evaluate regret (following the definitions in \eqref{reg_eq}) for both methods, our method significantly outperforms the MPC-based approach (as shown in Fig. \ref{comp_regret}). These results suggest that our approach offers improved robustness against parameter uncertainty, albeit at the expense of a marginal increase in total cost.       
\section{Concluding remarks and Future work}
In this paper, we introduce a game-theoretic horizon-splitting approach to model the conflicting objectives of financial gains from V2G participation and battery degradation from the user perspective. This methodology is a more robust alternative to existing approaches, offering system designers the ability to fine-tune the balance between V2G exploitability and battery health using adjustable hyperparameters. Our trade-off analysis yields the following key results: 1) For EVs with higher battery capacities, users should consider higher levels of V2G participation, as the relative impact on battery degradation is reduced. 2) Users are encouraged to increase V2G participation during periods of high volatility in V2G pricing, particularly when using high-power chargers, as this can maximize financial benefits with low impact on battery health. 

\textcolor{black}{Future research will focus on implementing our proposed methodology in real-world test scenarios. Particularly, using real V2G tariffs and measuring the capacity of the EV battery, we aim to validate our theoretical conclusions. Additionally, as V2G services are heavily dependent on electricity pricing aligned with grid demands, our next objective is to develop a fair pricing mechanism using learning-based approaches and integrate it into our smart charging framework.} %

\section{Appendix}
\subsection{Derivation of equation \eqref{loss_model}}
From \eqref{emp_model}, we write
\begin{align*}
      Q_{\text{loss},\%}^{\text{cyc}} & =B_{1} (\exp{(B_{2}|I_{\text{rate}}|)}C_{\text{rated}}n_{\text{cycle}}).
\end{align*}
As, we want the battery pack level power $P_{\text{bat}}$ to be brought down to cell level c-rate i.e., $I_{\text{rate}}$, considering a total number of $s= n_{\text{series}}\times n_{\text{parallel}}$ cells, rated battery voltage $V_{\text{bat}}$, and capacity $C_{\text{rated}}$,  
\begin{align*}
      Q_{\text{loss},\%}^{\text{cyc}} & =B_{1} (\exp{\left(\frac{B_2|P_{\text{bat}}|}{sV_{\text{bat}}C_{\text{rated}}}\right)}C_{\text{rated}}n_{\text{cycle}}),
\end{align*}
$n_{\text{cycle}}$ is the number of full charging-discharging cycles. We want to know the fraction of a cycle at every interval ($\Delta t$). For that, we approximate $n_{\text{cycle}}= \frac{n_{\max}\Delta t}{T}$ where $n_{\max}$ is the maximum number of possible full-cycles in a single V2G session of $T$ intervals. In addition, the term $Q^{\text{cyc}}_{\text{loss},\%}$ is the percentage capacity loss of rated capacity ($C_{\text{rated}}$). We find the capacity loss in Ah at time $t$ as, $Q^{\text{cyc}}_{\text{loss},t}= \frac{C_{rated}}{100}Q^{\text{cyc}}_{\text{loss},\%}$. Finally, by representing $\hat{n}= \frac{n_{\max}\Delta t}{T\times 100}$, we incorporate the above changes as, 
\begin{align*}
      Q_{\text{loss},t}^{\text{cyc}} & =B_{1,t} C^2_{\text{rated}}\hat{n} (\exp{\left(\frac{B_{2,t}|P_{\text{bat,t}}|}{sV_{\text{bat}}C_{\text{rated}}}\right)}),
\end{align*}
The above function is convex in $P_{\text{bat},t}$ but non-smooth at $P_{\text{bat},t}=0$. One could always use optimization solvers that can handle non-smooth convex optimization problems. However, we are interested in the smooth version of it as it becomes more computationally efficient, and standard solvers are easily available. Therefore, we consider the approximation of $\exp^{\alpha|x|} \approx (1+\frac{\alpha^2x^2}{h})$, where $h$ being a curve fitting parameter. After replacing the corresponding terms, we have 
\begin{align*}
    Q_{\text{loss},t}^{\text{cyc}} & = B_{1,t}C_{\text{rated}}^2 \hat{n} \left(1+\frac{\hat{B}_{2,t}^2P_{\text{bat},t}^2}{hs^2} \right), 
\end{align*}
where $\hat{B}_{2,t}= \frac{B_{2,t}}{V_{\text{bat}}C_{\text{rated}}}$.
\subsection{\textit{Proof of Proposition 1}} The local feasibility set of \eqref{obj_1} and \eqref{obj_2} are convex and there exists $\Omega := \prod_{i=1}^{2} \Omega_i(u^{-i}) $, which is a nonempty, closed and convex set. Therefore, our problem satisfies conditions of \textit{Definition $2$}. Now, we define the following exact potential function candidate: 
    \begin{align*}
        P(\mathbf{u}):= \sum_{j\in \mathcal{T}_m^w} (\alpha_j P_{\text{bat},j} \Delta t) + \gamma\sum_{j \in (\Tilde{\mathcal{T}}_m^w)} Q^{\text{cyc}}_{\text{loss},j}(P_{\text{bat},j}). 
    \end{align*}
    Given $ x^1, y^1 \in \Omega_1 (u^{-1})$, $x^1:=[\hat{P}_{\text{bat},t}], y^1:=[\Tilde{P}_{\text{bat},t}], \forall t \in \mathcal{T}^w_m$, and $u^{-1}:= [P_{\text{bat},t}], \forall t \in \Tilde{\mathcal{T}}_m^w$, for player $1$ (V2G), we have 
    \begin{align*}
        & P(x^1,u^{-1}) - P(y^1,u^{-1}) = \sum_{j \in \mathcal{T}^w_m} \alpha_j (\hat{P}_{\text{bat},j}- \Tilde{P}_{\text{bat},j}) \Delta t \\
        &= \left[\sum_{j \in \mathcal{T}^w_m} (\alpha_j \hat{P}_{\text{bat},j} \Delta t) + \sum_{j \in \Tilde{\mathcal{T}}_m^w} (\alpha_j P_{\text{bat},j} \Delta t)\right] \\
        & -\left[ \sum_{j \in \mathcal{T}^w_m} (\alpha_j \Tilde{P}_{\text{bat},j} \Delta t) + \sum_{j \in \Tilde{\mathcal{T}}_m^w} (\alpha_j P_{\text{bat},j} \Delta t) \right] \\
        &= \theta^1(x^1,u^{-1})- \theta^1(y^1,u^{-1}) \quad (\text{in view of \eqref{obj_1}}). 
    \end{align*}
    Similarly, given $x^2, y^2 \in \Omega_1 (u^{-2})$, $x^2:=[\hat{P}_{\text{bat},t}], y^2:=[\Tilde{P}_{\text{bat},t}], \forall t \in \Tilde{\mathcal{T}}_m^w$, and $u^{-2}:= [P_{\text{bat},t}], \forall t \in \mathcal{T}^w_m$, for player $2$ (BD), we have 
    \begin{align*}
        &P(x^2,u^{-2}) - P(y^2,u^{-2})  \\
        &= \sum_{j \in \Tilde{\mathcal{T}}_m^w} \gamma \left(Q^{\text{cyc}}_{\text{loss},j}(\hat{P}_{\text{bat},j})- Q^{\text{cyc}}_{\text{loss},j}(\Tilde{P}_{\text{bat},j})\right) \\
        &= \gamma\left[ \sum_{j \in \Tilde{\mathcal{T}}_m^w}      
         Q^{\text{cyc}}_{\text{loss},j}(\hat{P}_{\text{bat},j}) + \sum_{j \in \mathcal{T}^w_m}  Q^{\text{cyc}}_{\text{loss},j}(P_{\text{bat},j}) \right]  \\ 
         &- \gamma\left[ \sum_{j \in \Tilde{\mathcal{T}}_m^w}      
         Q^{\text{cyc}}_{\text{loss},j}(\Tilde{P}_{\text{bat},j}) + \sum_{j \in \mathcal{T}^w_m}  Q^{\text{cyc}}_{\text{loss},j}(P_{\text{bat},j}) \right] \\
         &=  \theta^2(x^2,u^{-2})- \theta^2(y^2,u^{-2}) \quad (\text{in view of \eqref{obj_2}} ).
    \end{align*}
    Therefore, $P(\textbf{u})$ is an exact potential function according to \textit{Definition $2$}. \hfill $\blacksquare$

\bibliographystyle{IEEEtran}
\bibliography{references.bib}
	
\end{document}